\documentclass[12pt, draftclsnofoot, onecolumn]{IEEEtran} % regular single-column

\usepackage{xspace,amsmath,amssymb,amsfonts,epsfig,syntonly,subfigure}
\usepackage{cite,bm,color,url,textcomp,empheq,boxedminipage}
\usepackage{algorithmicx,algorithm,algpseudocode}
\usepackage{epstopdf,makecell}
\usepackage{empheq}
\usepackage{graphicx,graphics}  % Written by David Carlisle and Sebastian Rahtz
\usepackage{subfigure}
\usepackage{multirow,multicol}
\usepackage{psfrag}    % Written by Craig Barratt, Michael C. Grant,
\usepackage{stfloats}
\usepackage{url}
\usepackage{array}
\usepackage{booktabs} %
\usepackage{multirow}
\usepackage{CJK}
\usepackage{times}
\usepackage{cite,graphicx,amsmath,amssymb}
\usepackage{subfigure}
\usepackage{fancyhdr}
\usepackage{hhline}
\usepackage{graphicx,graphics}
\usepackage{array,color}
\usepackage[dvipsnames, svgnames, x11names]{xcolor}
\usepackage{amsmath}
\usepackage{amsthm}

\newtheorem{proposition}{Proposition}

\newtheorem{lemma}{Lemma}

\newcommand{\mv}[1]{\mbox{\boldmath{$ #1 $}}}

\PassOptionsToPackage{bookmarks={false}}{hyperref}

\include{header}

\long\def\symbolfootnote[#1]#2{\begingroup
\def\thefootnote{\fnsymbol{footnote}}
\footnote[#1]{#2}\endgroup}

\psfull

\IEEEoverridecommandlockouts

\begin{document}

\title{Joint Communication and Computation Optimization for Wireless Powered Mobile Edge Computing with D2D Offloading}
\author{Dixiao Wu, Feng Wang, Xiaowen Cao, and Jie Xu\\
\thanks{Part of this paper has been presented at the IEEE Global Communications Conference (Globecom) Workshop on Wireless Energy Harvesting Communication Networks, Abu Dhabi, United Arab Emirates, Dec. 9--13, 2018 \cite{Conf_version}.}

\thanks{The authors are with the School of Information Engineering, Guangdong University of Technology, Guangzhou 510006, China (e-mail: 3214003073@mail2.gdut.edu.cn, fengwang13@gdut.edu.cn, 2111703002@mail2.gdut.edu.cn, jiexu@gdut.edu.cn). J. Xu is the corresponding author.}
}

%\markboth{}{}
\maketitle

\begin{abstract}
This paper studies a wireless powered mobile edge computing (MEC) system with device-to-device (D2D)-enabled task offloading. In this system, a set of distributed multi-antenna energy transmitters (ETs) use collaborative energy beamforming to wirelessly charge multiple users. By using the harvested energy, the actively computing user nodes can offload their computation tasks to nearby idle users (as helper nodes) via D2D communication links for self-sustainable remote computing.
We consider the frequency division multiple access (FDMA) protocol, such that the D2D communications of different user-helper pairs are implemented over orthogonal frequency bands.
Furthermore, we focus on a particular time block for task execution, which is divided into three slots for computation task offloading, remote computing, and result downloading, respectively, at different user-helper pairs.
Under this setup, we jointly optimize the collaborative energy beamforming at ETs, the communication and computation resource allocation at users and helpers, and the user-helper pairing, so as to maximize the sum computation rate (i.e., the number of task input-bits executed over this block) of the users, subject to individual energy neutrality constraints at both users and helpers.
First, we consider the computation rate maximization problem under any given user-helper pairs, for which an efficient solution is proposed by using the techniques of alternating optimization and convex optimization.
Next, we develop the optimal user-helper pairing scheme based on exhaustive search and a low-complexity scheme based on greedy selection.
Numerical results show that the proposed design significantly improves the sum computation rate at users, as compared to benchmark schemes without such joint optimization.
\end{abstract}

%Note that keywords are not normally used for peerreview papers.
%\newpage
%\IEEEpeerreviewmaketitle

%\bibliography{mec_May27}
%\bibliographystyle{IEEEtran}
\IEEEpeerreviewmaketitle

\section{Introduction}
With recent advancements in artificial intelligence (AI), big data, and Internet of things (IoT), it is envisioned that future wireless networks need to support massive low-power wireless devices (e.g., sensors and actuators) with real-time communication and computation, in order to enable various new applications such as industrial automation, smart transportation, and unmanned aerial vehicles (UAVs). Towards this end, how to provide rich computation capability and sustainable energy supply for these wireless devices is becoming a critical technical challenge to be tackled.

Recently, mobile edge computing (MEC) has emerged as a promising solution to provide wireless devices with cloud-like computing at the network edge (such as base stations (BSs) and access points (APs)) \cite{1,2,3}. Compared to the conventional cloud computing with centralized computing centers that are far away, MEC enables wireless devices to offload their computation tasks to nearby edge servers distributed at BSs/APs for remote execution. This thus helps these devices enjoy significantly reduced computation latency and computation/communication reliability, and relieves the unpredictable traffic congestion in backbone networks. On the other hand, radio-frequency (RF) signal based wireless power transfer (WPT) has also been recognized as a viable and cost-efficient solution to wirelessly charge a large number of low-power electronic devices in future wireless networks, in which dedicated energy transmitters (ETs) are deployed for wireless energy broadcasting (see \cite{4,general,5,6,7} and references therein).
Simultaneous wireless information and power transfer (SWIPT) and wireless powered communication networks (WPCNs) are two main WPT paradigms in wireless communications, which aim to provide sustainable wireless communications in the IoT era \cite{8,10,lifeng,bisuzhi,khuang,mxia}.

To exploit both benefits of MEC and WPT, wireless powered MEC has been recently proposed to achieve self-sustainable computing for wireless devices, in which a new type of hybrid APs are employed to serve as both ETs and MEC servers. These hybrid APs not only wirelessly charge these devices via WPT, but also help remotely execute their offloaded computation tasks \cite{12,13,14,15}.
In the literature, the prior work \cite{12} first considered a wireless powered single-user MEC system, in which the user's successful computation probability under given computation latency is maximized, by jointly optimizing the WPT at the AP as well as the local computing and computation offloading at the user. The work \cite{13} further investigated the wireless powered multiuser MEC system under a time-division multiple access (TDMA) based partial offloading protocol, in which the communication and computation resource allocations are jointly optimized to minimize the total system energy consumption under the users' computation latency constraints. Furthermore, the authors in \cite{14} studied a wireless powered multiuser MEC system with inseparable (or binary) computation tasks at each user, and \cite{15} investigated a wireless powered relaying system for efficient MEC.

Despite the research efforts on wireless powered MEC, the above works \cite{12,13,14,15} focused on the scenario with the edge server co-located at the hybrid AP with adequately large computation capacity.
Such a design, however, may not be applicable in other WPT scenarios when only ETs are dedicatedly deployed without computation capabilities. Also, this design fails to exploit the rich computation resources at surrounding end users (such as smart IoT devices that are densely deployed in wireless networks nowadays). Due to the burst nature of wireless traffic, it is highly likely that, when some devices are actively computing, there exist some surrounding idle devices with unused computation resources, where all the devices can opportunistically harvest wireless energy from ETs based on the broadcast nature of RF energy signals. Motivated by this fact together with the recent success of device-to-device (D2D) communications \cite{wuqingqing, fengdaquan}, we propose a new wireless powered D2D offloading approach to exploit both the unused computation resources and the opportunistic wireless energy harvesting at surrounding idle devices. In this approach, these idle devices are enabled as friendly helpers to contribute their opportunistically harvested wireless energy to help remotely execute the computation tasks offloaded from active users via D2D links, thereby improving the computation performance. Notice that the D2D-enabled cooperative computation has been investigated in prior works \cite{xiaowen,qijieLin,hongXing,pulingjun,chenxu}. However, these works only considered the users' cooperative computation under fixed energy supplies (e.g., batteries). By contrast, this paper unifies both cooperative computing and WPT, where the energy consumption of helpers comes from the wireless energy transferred from the ET, thus leading to self-sustainable computation cooperation among users.

%Nevertheless, there still exist some challenges waiting to be tackled. Unlike the above works who just considered the cooperative computation at users enabled by themselves' limited energy batteries, a more interesting and sustainable joint design that combines WPT and cooperative computation at the users is proposed in this paper to improve the computation performance of MEC system. However, based on energy harvesting,  the communication and computation resources allocation would be more challenging, while due to the coupling of both offloading and downloading at users for the cooperative computation, the time allocation under the delay constraint at users would further increase the difficulty of the proposed joint design.
%Besides, by allowing multiple users accessing in this system, establishing an efficient pairing strategy among users to enable the cooperative communication is also extremely urgent. this paper unifies both cooperative computing and collaborative energy (beamforming), where the energy consumption of helpers comes from the wireless energy transferred from the ETs, thus leading to self-sustainable computation cooperation among users.
In particular, we consider a wireless powered MEC system with D2D-enabled task offloading, where a set of distributed multi-antenna ETs use collaborative energy beamforming to wirelessly charge multiple users and helpers. By using the harvested energy, the actively computing users offload their computation tasks to nearby helpers via D2D communication links for self-sustainable remote computing.
For ease of analysis and idea exposure, we consider that the WPT and the D2D offloading are implemented over orthogonal frequency bands, and the D2D communications of different user-helper pairs are operated following the frequency division multiple access (FDMA) protocol.
Furthermore, we focus on a particular time block for task execution, which is divided into three slots for computation task offloading, remote computing, and result downloading, respectively, for all the user-helper pairs. Under this setup, we jointly optimize the collaborative energy beamforming at the $N$ ETs, the communication and computation resource allocations at both the users and helpers, and the user-helper pairing policy, so as to maximize the sum computation rate (i.e., the number of task input-bits executed over this block) of the users, subject to the individual transmit power constraints at the ETs, and the individual energy neutrality constraints at both the users and helpers.
Due to the combinatorial nature of user-helper paring and the coupling of the time and frequency allocation variables, the formulated problem is highly non-convex and challenging to be optimally solved.
To tackle this issue, we first consider the computation rate maximization problem under any given user-helper pairs, for which an efficient solution is proposed based on the techniques of alternating optimization and convex optimization. Next, we develop the optimal user-helper pairing scheme based on exhaustive search and a low-complexity pairing scheme based on the greedy-based selection.
Numerical results are provided to show that the proposed designs significantly improve the sum computation rate at users, as compared to benchmark schemes without such joint optimization.

The remainder of this paper is organized as follows. Section~\ref{sec:system} introduces the system model and formulates the sum computation rate maximization problem. Section~\ref{sec:design} presents an efficient solution to the formulated problem under given user-helper pairing. Section~\ref{sec:pairing} presents various user-helper pairing schemes. Section~\ref{sec:results} provides numerical results, followed by the conclusion in Section~\ref{sec:conclusion}.

{\em Notations}: Boldface letters refer to vectors (lower case) or matrices (upper case). $\boldsymbol{I}$ and $\boldsymbol{0}$ denote an identity matrix and an all-zero vector/matrix, respectively. For a square matrix $\boldsymbol{M}$, ${\rm tr}(\boldsymbol{M})$ denotes its trace, and $\boldsymbol{M} \succeq \mv 0$ means that $\boldsymbol{M}$ is positive semidefinite. For an arbitrary-size matrix $\boldsymbol{A}$, $\boldsymbol{A}^\dagger$ denotes its transpose and $\boldsymbol{A}^H$ denotes its conjugate transpose.
$\mathbb {C}^{x\times y}$ denotes the space of $x\times y$ complex matrices; $\mathbb {R}$ denotes the set of real numbers; $\mathbb {E}[\cdot]$ denotes the statistical expectation. $\|\boldsymbol {x}\|$ denotes the Euclidean norm of a vector $\boldsymbol {x}$, and $[x]^{+} \triangleq \max (x,0)$.

\section{System Model and Problem Formulation}\label{sec:system}

\begin{figure}
 \centering
 \includegraphics[width=6.0in]{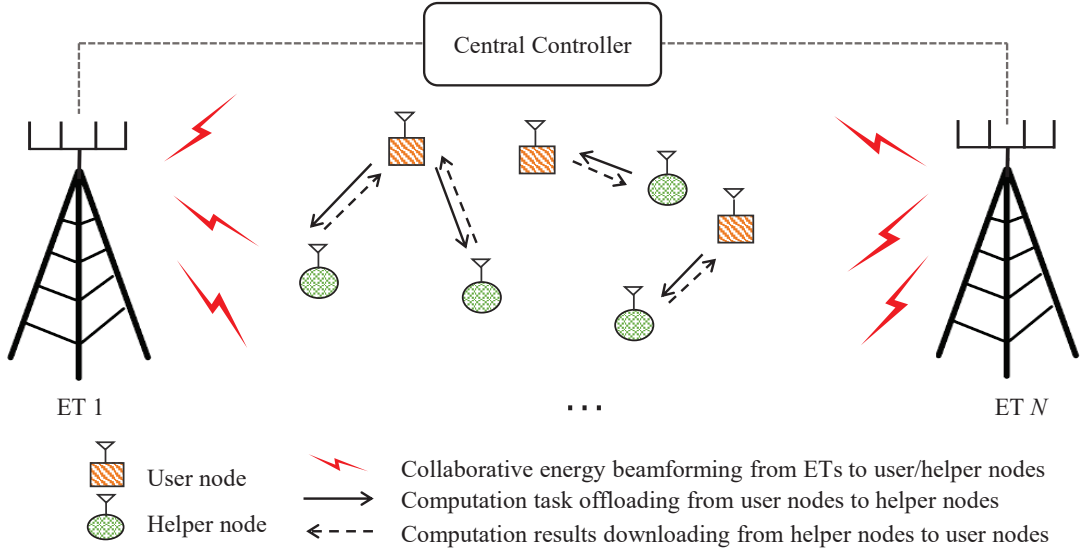}
 \caption{ System model of a wireless powered MEC system with D2D-enabled computation offloading.}\label{fig:SysMod}
\end{figure}

We consider a wireless powered D2D-enabled cooperative computing system as shown in Fig.~\ref{fig:SysMod}, which consists of a set ${\cal N}\triangleq \{1,\ldots,N\}$ of ETs, a set ${\cal K} \triangleq \{1,\ldots,K\}$ of users, and a set ${\cal M} \triangleq \{1,\ldots,M\}$ of helpers.
It is assumed that each ET is equipped with $N_t \ge 1$ antennas and each user or helper node is equipped with one single antenna. In this system, both the user and helper nodes are wirelessly energized by the WPT from the ETs. Relying on the harvested energy, each user $k\in{\cal K}$ executes the computation tasks by local computing and offloading to the paired helpers for cooperative computing.
Without loss generality, we consider that each helper $m \in \mathcal M$ can only be paired with one user for computation cooperation. Let ${\cal M}_k \triangleq \{1,\ldots,M_k\}$ denote the set of helpers that are paired with user $k$, $\forall k\in\cal K$. Then, we have $ {\cal M}_i \cap {\cal M}_j = \phi$, $\forall i, j\in {\cal{K}}, i\neq j$, and $\cup_{k\in{\cal K}} {\cal M}_k \subseteq {\cal M} $.
In addition, to avoid the co-channel interference, we assume that the WPT from the ETs to users/helpers and the computation offloading between users and their paired helpers are operated over orthogonal frequency bands. We focus on one particular time block with duration $T$, during which the wireless channels are assumed to remain unchanged. Suppose that each user has latency-constrained computation tasks to be executed, whose deadline should not exceed the duration $T$. Furthermore, we assume that there exists a central controller that can collect the global channel state information (CSI) and computation-related information and can thus accordingly coordinate the WPT and D2D offloading at the users and helpers.

\subsection{Collaborative Energy Beamforming at ETs}
First, we consider the collaborative energy beamforming from the $N$ ETs to the $K$ users and $M$ helpers. Let $\bm s_n \in\mathbb{C}^{N_t\times 1} $ denote the energy-bearing signal sent by ET $n \in \cal N$, and ${\bm s} \triangleq \left[{\bm s}_1^\dagger,\ldots,{\bm s}_N^\dagger\right]^\dagger$ denote the vector collecting the energy signals at the $N$ ETs. Suppose that ${\bm s}$ is generated based on pseudo-random sequences that are known {\it a-priori} by all the ETs. Accordingly, we define ${\bm S} \triangleq \mathbb{E}\left[{\bm s}{\bm s}^H\right] \in \mathbb{C}^ {N_t N \times N_t N}$ as the transmit energy covariance matrix to be optimized later. Accordingly, the transmit power at ET $n\in{\cal N}$ is given by $\mathbb{E}\left[||\bm s_n||^2\right] = {\rm tr}\left( \mv \Sigma_n \bm S \right)$, where $\mv \Sigma_n \in \mathbb{C}^{N_tN\times N_tN}$ is a diagonal matrix with its $\left((n-1)N_t+1\right)$-th to $nN_t$-th diagonal elements being one and zero otherwise. Suppose that the transmit power budget at each ET $n\in \mathcal N$ is $P_n$. Then we have
\begin{align}\label{power}
{\rm tr}( \mv \Sigma_n \bm S ) \leq P_{n}, \forall n\in \mathcal N.
\end{align}
Let ${\bm g}_{k,0,n} \in\mathbb{C}^{N_t\times 1}$ and ${\bm g}_{k,m,n} \in\mathbb{C}^{N_t\times 1}$ denote the channel vectors from ET $n \in \mathcal N$ to user $k\in{\cal K}$ and helper $m \in {\cal M}_k$ (paired with user $k$), respectively. Define ${\bm g}_{k,0}\triangleq \left[{\bm g}^\dagger_{k,0,1},\ldots,{\bm g}^\dagger_{k,0,N}\right]^\dagger \in\mathbb{C}^{N_tN\times 1}$ and ${\bm g}_{k,m}\triangleq \left[{\bm g}^\dagger_{k,m,1},\ldots,{\bm g}^\dagger_{k,m,N}\right]^\dagger \in \mathbb{C}^{N_tN\times 1}$ as the combined channel vectors from all $N$ ETs to user $k \in \cal {K}$ and helper $m\in {\cal M}_k$, respectively. The amounts of energy harvested by user $k$ and helper $m$ during the block duration $T$ are respectively given by
\begin{align}
&E_{k,0} = T\eta\mathbb{E}\left[|{\bm g}_{k,0}^H \bm s|^2\right] =T \eta{\rm tr}\left( {\bm g}_{k,0}^H\bm S {\bm g}_{k,0}\right), ~\forall k\in \mathcal K,\\
&E_{k,m} = T\eta\mathbb{E}\left[|{\bm g}_{k,m}^H \bm s|^2\right] = T \eta{\rm tr}\left( {\bm g}_{k,m}^H\bm S {\bm g}_{k,m}\right),~ \forall m\in \mathcal M_k, ~k\in \mathcal K,
\end{align}
where $0 < \eta \leq 1$ denotes the constant energy conversion efficiency.

\begin{figure}
 \centering
 \includegraphics[width=5.0in]{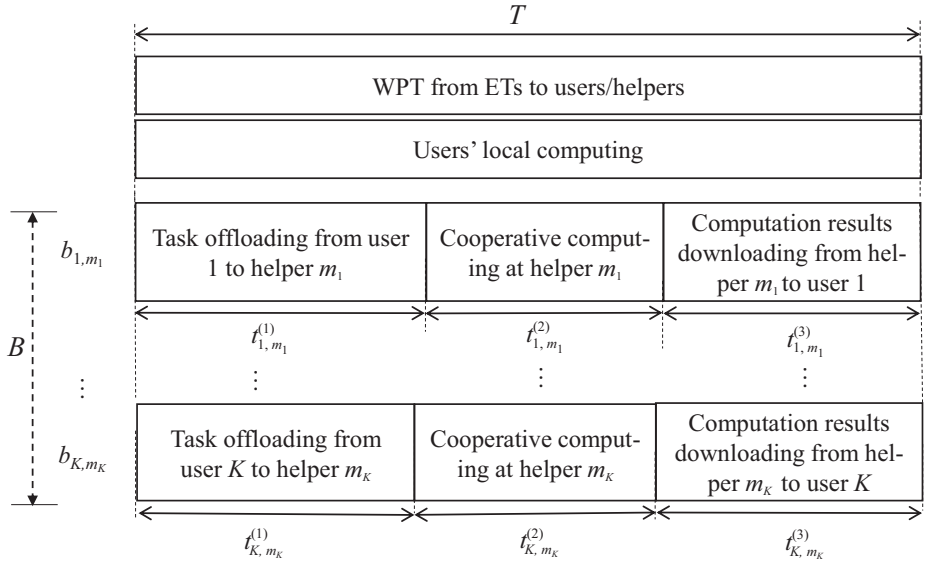}
 \caption{ An illustration of the TDMA-based computation offloading protocol, and each user-helper pair's D2D offloading and WPT are operated over orthogonal frequency bands. }\label{fig:protocol}
\end{figure}

\subsection{D2D-Enabled Cooperative Computing}
We consider the partial offloading at the users, such that each user $k\in{\cal K}$ can arbitrarily partition the computation tasks into several parts for computing locally and offloading to its paired helpers, respectively. Denote $L_k$ as the number of task input-bits at each user $k\in{\cal K}$. To proceed, these task input-bits at user $k$ can be partitioned into $|{\cal M}_k |+1$ parts, such that user $k$ computes $\ell_{k,0}$ task input-bits locally and offloads $\ell_{k,m}$ task input-bits to each helper $m\in{\cal M}_k$ for remote execution simultaneously. We thus have $L_k = \ell_{k,0}+\sum_{m\in {\cal M}_k} \ell_{k,m}$. We consider the TDMA-based offloading protocol as shown in Fig. \ref{fig:protocol}, where the D2D-enabled task offloading and result downloading between user $k$ and each paired helper $m \in {\cal M}_k$ (represented by $m_k$ in Fig. \ref{fig:protocol}) is allocated with an orthogonal frequency band with bandwidth $b_{k,m}$. Accordingly, we obtain the bandwidth allocation constraint as
\begin{align}\label{frequence}
 \sum_{k \in \cal K}\sum_{m\in {\cal M}_k}b_{k,m} \leq B.
\end{align}
Furthermore, regarding the cooperative computation, the time block with duration $T$ is divided into three slots with duration $t_{k,m}^{(1)}$, $t_{k,m}^{(2)}$, and $t_{k,m}^{(3)}$ (to be optimized later), for offloading from user $k$ to helper $m \in {\cal M}_k$, cooperative computing at helper $m \in {\cal M}_k$, and results downloading from helper $m \in {\cal M}_k$ to user $k$, respectively. As a result, we have the time allocation constraints as
\begin{align}\label{duration}
\sum_{i=1}^3 t_{k,m}^{(i)} \leq T, ~\forall m \in {\cal M}_k, ~k\in\cal K.
\end{align}
In the following, we explain the D2D-enabled computation offloading between the users and helpers in the three consecutive slots in details, respectively.

\subsubsection{Task Offloading from User $k$ to Helper $m \in {\cal M}_k$}
In the first slot, user $k$ offloads the computation tasks with $\ell_{k,m}$ task input-bits to helper $m \in {\cal M}_k$ over an orthogonal channel with bandwidth $b_{k,m}$. Let $h_{k,m}>0$ denote the channel power gain between user $k$ and helper $m \in {\cal M}_k$, and $q_{k,m}$ denote the transmit power at user $k$ for task offloading to helper $m \in {\cal M}_k$. Accordingly, the number of offloaded task input-bits from user $k$ to helper $m \in {\cal M}_k$ is given by
\begin{align}\label{l_hm}
\ell_{k,m}=t_{k,m}^{(1)}b_{k,m} \log_2\left(1+\frac{h_{k,m}q_{k,m}}{N_0b_{k,m}}\right),
\end{align}
where $N_0$ denotes the power spectral density (PSD) of the additive white Gaussian noise (AWGN) at the user/helper receivers. As a result, by combining the offloading energy consumption towards all the helpers in ${\cal M}_k$, the total communication energy consumption by user $k\in\cal K$ is given by
\begin{align}
E_{k,0}^{\rm tx} = \sum_{m \in {\cal M}_k} q_{k,m}t_{k,m}^{(1)} = \sum_{m \in {\cal M}_k} \frac{N_0t_{k,m}^{(1)}b_{k,m}}{h_{k,m}}\left(2^{\frac{\ell_{k,m}}{t_{k,m}^{(1)}b_{k,m}}}-1\right).
\end{align}

\subsubsection{Cooperative Computation at Helper $m \in {\cal M}_k$}
In the second slot with duration $t_{k,m}^{(2)}$, each helper $m \in {\cal M}_k$ needs to execute the offloaded task with the number of $\ell_{k,m}$ task input-bits for cooperative computation.
Let $C_{k,m}$ denote the central process unit (CPU) cycles for executing one task input-bit at helper $m \in {\cal M}_k$. Accordingly, a total of $C_{k,m}\ell_{k,m}$ CPU cycles are required for helper $m \in {\cal M}_k$ to accomplish the task execution. In this case, the energy consumption of helper $m \in {\cal M}_k$ is expressed as  $E_{k,m}^{\rm comp}=\sum_{j=1}^{C_{k,m}\ell_{k,m}} \xi_{k,m} f_{k,m,j}^2$ \cite{votage}, where $\xi_{k,m}$ and $f_{k,m,j}$ denote the effective CPU switch capacitance and CPU frequency for executing the $j$-th CPU cycle at helper $m \in {\cal M}_k$, respectively. To minimize the energy consumption of helper $m \in {\cal M}_k$ for remote computing, an identical CPU frequency should be adopted for every CPU cycle, i.e., we have $f_{k,m,1}=f_{k,m,2}=\ldots=f_{k,m,C_{k,m}\ell_{k,m}}={C_{k,m}\ell_{k,m}}/{t_{k,m}^{(2)}}$ \cite{3}.
Correspondingly, the energy consumption for cooperative computing at helper $m \in {\cal M}_k$ is given as
\begin{align}
E_{k,m}^{\rm comp} = \frac{\xi_{k,m}C_{k,m}^3\ell_{k,m}^3}{\left(t_{k,m}^{(2)}\right)^2}.
\end{align}

\subsubsection{Computation Results Downloading from Helper $m \in {\cal M}_k$ to User $k$}
In the third slot with duration $t_{k,m}^{(3)}$, user $k\in\cal K$ downloads the corresponding computation results from each paired helper $m\in{\cal M}_k$. It is assumed that the size of the computation results is given by $\beta \ell_{k,m}$, which is proportional to that of offloaded task input-bits, and the system parameter $\beta$ depends on the task application type \cite{3}. Let $p_{k,m}$ denote the transmit power of helper $m \in {\cal M}_k$ for sending results back to user $k$. It thus follows that
\begin{align}
\beta\ell_{k,m} = t_{k,m}^{(3)}b_{k,m} \log_2\left(1+\frac{h_{k,m}p_{k,m}}{N_0b_{k,m}}\right),~\forall m\in {\cal M}_k, ~k\in{\cal K}.
\end{align}
As a result, the communication energy consumption at helper $m \in {\cal M}_k$ is given by
\begin{align}
E_{k,m}^{\rm tx} = p_{k,m} t_{k,m}^{(3)}=\frac{N_0t_{k,m}^{(3)}b_{k,m}}{h_{k,m}}\left(2^{\frac{\beta \ell_{k,m}}{t_{k,m}^{(3)}b_{k,m}}}-1\right).
\end{align}

\subsection{Local Computing at Users}
Note that each user $k\in\cal K$ executes its computation tasks with $\ell_{k,0}$ task input-bits within the whole block. Let $C_{k,0}$ denote the number of CPU cycles required for executing each task input-bit at user $k\in\cal K$. Similarly, an identical CPU frequency $C_{k,0}\ell_{k,0}/T$ is adopted for each CPU cycle at user $k\in\cal K$. The resultant energy consumption for user $k$'s local computing is given by \cite{3}
\begin{align}
E_{k,0}^{\rm comp} = \frac{\xi_{k,0}C_{k,0}^3\ell_{k,0}^3}{T^2},
\end{align}
where $\xi_{k,0}$ is the constant switch capacitance of CPU architecture at  user $k\in\cal K$.

\subsection{Problem Formulation}
Notice that in order to maintain the self-sustainable operation of the wireless powered MEC system, all users and helpers should only rely on the energy harvested from the ETs to complete their computation tasks. Therefore, the so-called energy neutrality constraints \cite{7} are imposed at both the users and helpers, such that within the block $T$, the total energy consumed by each user or helper node can not exceed the harvested energy from the ETs, i.e.,
%\begin{subequations}\label{energy consumption}
\begin{align}
& E_{k,0}^{\rm comp} +E_{k,0}^{\rm tx} \leq T \eta{\rm tr}( {\bm g}_{k,0}^H\bm S {\bm g}_{k,0}), ~\forall k\in{\cal K},\label{energy consumption_users}\\
& E_{k,m}^{\rm comp}+E_{k,m}^{\rm tx}\leq T \eta{\rm tr}( {\bm g}_{k,m}^H\bm S {\bm g}_{k,m}), ~\forall m\in {\cal M}_k, ~k\in{\cal K}.\label{energy consumption_helpers}
\end{align}
%\end{subequations}

In this paper, we aim to maximize the sum computation rate (i.e., the total number of task input-bits executed within the block) at the $K$ users, by jointly optimizing the user-helper pairing sets $\{{\cal{M}}_k\}$, the energy transmit covariance matrix $ \bm S$ for collaborative energy beamforming at ETs, the task allocation $\{\ell_{k,m}\}$, the time allocation $\{t_{k,m}^{(1)}, t_{k,m}^{(2)}, t_{k,m}^{(3)}\}$, and bandwidth allocation $\{b_{k,m}\}$. Mathematically, the users' sum computation rate maximization problem is formulated as
\begin{subequations}\label{eq.prob1}
\begin{align}
\mathbf{(P1):} \max_{
 \{{\cal M}_k, \ell_{k,0}, \ell_{k,m}, b_{k,m}, t_{k,m}^{(i)}\},\bm S}~&  \sum_{k \in \cal K} \left(\ell_{k,0}+\sum_{m\in {\cal M}_k} \ell_{k,m}\right)\notag\\
{\text {s.t.}}~~~~~~~~~~~~&\eqref{power},~\eqref{frequence}, ~\eqref{duration} ,~\eqref{energy consumption_users}, ~{\text {and}} ~\eqref{energy consumption_helpers}\notag\\
&\bigcup_{k\in{\cal{K}}}{\cal{M}}_k\subseteq {\cal{M}}, ~{\cal{M}}_i\cap{\cal{M}}_j=\phi,~\forall i, j \in {\cal{K}},~i \neq j\label{eq.prob_M}\\
&t_{k,m}^{(i)}\geq 0, ~\forall i\in\{1,2,3\},~ m\in {\cal M}_k,~  k\in \cal K\label{eq.prob1_t}\\
& \ell_{k,0}\geq 0,~\ell_{k,m}\geq 0,~\forall m\in {\cal M}_k, ~ k\in{\cal K}\label{eq.prob1_l}\\
& b_{k,m}\geq 0,~\forall m\in {\cal M}_k, ~ k\in{\cal K}\label{eq.prob1_b}\\
& \bm S \succeq \bm 0. \label{eq.prob1_S}
\end{align}
\end{subequations}
Note that problem (P1) is non-convex due to the involvement of user-helper pairing sets $\{{\cal{M}}_k$\} and the coupling between $\{t_{k,m}^{(i)}\}$ and $\{b_{k,m}\}$. Therefore, problem (P1) is difficult to be optimally solved. In the following, we first present an efficient algorithm to solve problem (P1) under any given user-helper pairing sets $\{{\cal M}_k\}$ in Section \ref{sec:design}, and then develop algorithms to find effective user-helper pairing designs in Section \ref{sec:pairing}.

\section{Proposed Solution to Problem (P1) under Given User-Helper Pairing}\label{sec:design}
In this section, we consider the computation rate maximization problem under any  given user-helper pairing sets $\{ {\cal M}_k \}$. In this case, problem (P1) is reduced as
\begin{align*}
\mathbf{(P2):}~\psi({\{\cal{M}}_k\}) = \max_{
\{\ell_{k,0}, \ell_{k,m}, b_{k,m}, t_{k,m}^{(i)}\},\bm S}~&  \sum_{k \in \cal K} \left(\ell_{k,0}+\sum_{m\in {\cal M}_k} \ell_{k,m}\right)\\
{\text {s.t.}}~~~~~~\eqref{power},&~ \eqref{frequence}, ~\eqref{duration} ,~\eqref{energy consumption_users}, ~\eqref{energy consumption_helpers},~{\text{and}} ~\eqref{eq.prob1_t}-\eqref{eq.prob1_S},\notag
\end{align*}
where $\psi(\{{\cal M}_k\})$ denotes the obtained sum computation rate under given $\{{\cal M}_k\}$. Due to the coupling of the time allocation $\{t_{k,m}^{(i)}\}$ and bandwidth allocation $\{b_{k,m}\}$ in constraints \eqref{energy consumption_users} and \eqref{energy consumption_helpers}, problem (P2) is still a non-convex optimization problem. Nonetheless, it is verified that problem (P2) becomes a convex optimization problem under either given time allocation $\{t_{k,m}^{(i)}\}$ or bandwidth allocation $\{b_{k,m}\}$. As presented in the following, an alternating-optimization-based approach can then be employed to solve problem (P2), in which we optimize over $\{\ell_{k,0}, \ell_{k,m}, t_{k,m}^{(i)}, \bm S\}$, as well as $\{\ell_{k,0}, \ell_{k,m}, b_{k,m}, \bm S\}$, via leveraging convex optimization techniques in an alternating manner, by considering $ \{b_{k,m}\}$ and $\{t_{k,m}^{(i)}\}$ to be given, respectively.

\subsection{Optimization over $ \{\ell_{k,0}, \ell_{k,m}, t_{k,m}^{(i)}, \bm S\}$ under Given $\{b_{k,m}\}$}
In this subsection, we consider problem (P2) under given bandwidth allocation $\{b_{k,m}\}$. In this case, problem (P2) is further reduced to the following convex optimization problem
\begin{align*}
\mathbf{(P2.1):} ~\psi_1(\{{\cal{M}}_k, b_{k,m}\}) = \max_{\{\ell_{k,0}, \ell_{k,m}, t_{k,m}^{(i)}\}, \bm S}~  \sum_{k \in \cal K}& \left(\ell_{k,0}+\sum_{m\in {\cal M}_k} \ell_{k,m}\right)\\
 {\text {s.t.}}~~~~~~ \eqref{power}, ~&\eqref{duration} ,~\eqref{energy consumption_users}, ~\eqref{energy consumption_helpers} ,~\eqref{eq.prob1_t}, ~\eqref{eq.prob1_l},~{\text {and}}~\eqref{eq.prob1_S},
\end{align*}
where $\psi_1(\{{\cal M}_k, b_{k,m}\})$ denotes the obtained sum computation rate under given $\{{\cal M}_k\}$ and $\{b_{k,m}\}$. To reveal the essential engineering insight, we use the Lagrange duality method to obtain the optimal solution to problem (P2.1). Let $\gamma_n \geq 0$, $ \rho_{k,m}\geq 0$,
$\lambda_{k} \geq 0 $, and $\mu_{k,m} \geq 0$, $\forall n \in \mathcal{N}$, $m \in {\cal M}_k$, $k \in \cal K$, denote the Lagrange multipliers associated with the constraints in \eqref{power}, \eqref{duration}, \eqref{energy consumption_users}, and \eqref{energy consumption_helpers}, respectively. Then, the partial Lagrangian of problem (P2.1) is expressed as
\begin{align*}
&{\cal L}(\bm S, \{\ell_{k,0}, \ell_{k,m},  t_{k,m}^{(i)}, \lambda_{k}, \mu_{k,m},  \rho_{k,m}, \gamma_n\})\\
%=  &\sum_{k \in \cal K}\left(\ell_{k,0}+\sum_{m\in {\cal M}_k} \ell_{k,m}\right)+ \sum_{k \in \cal K}\sum_{m \in {\cal M}_k} \rho_{k,m}\left(T-\sum_{i=1}^3 t_{k,m}^{(i)}\right)+ \sum_{n=1}^{N}\gamma_n\left(P_{n}-{\rm tr}( \mv \Sigma_n \bm S  )\right)\\
%+ &\sum_{k \in \cal K}\lambda_k \left(T\eta{\rm tr}({\bm g}_{k,m}^H \bm S{\bm g}_{k,m})- \sum_{m\in {\cal M}_k}\frac{N_0b_{k,m}t_{k,m}^{(1)}} {h_{k,m}}\left(2^{\frac{\ell_{k,m}}{t_{m,1}b_{k,m}}}-1\right) -\frac{\xi_{k,0}C_{k,0}^3\ell_{k,0}^3}{T^2}\right) \\
%+ &\sum_{k\in\cal{K}}\sum_{m\in {\cal M}_k} \mu_{k,m} \left(T \eta{\rm tr}\left({\bm g}_{k,m}^H \bm S{\bm g}_{k,m}\right)- \frac{N_0b_{k,m}t_{k,m}^{(3)}}{h_{k,m}}\left(2^{\frac{\beta\ell_{k,m}}{t_{k,m}^{(3)}b_{k,m}}}-1\right)
%+\frac{\xi_{k,m}C_{k,m}^3\ell_{k,m}^3}{(t_{k,m}^{(2)})^2}\right),\\
&= {\rm tr}\left( \left( \sum_{k \in \cal K}\left(\lambda_kT\eta{\bm g}_{k,0}{\bm g}_{k,0}^{H}+\sum_{m \in {\cal M}_k}\mu_{k,m}T\eta{\bm g}_{k,m}{\bm g}_{k,m}^{H}\right)-\sum\limits_{n\in\cal N}\gamma_n {\bm \Sigma_n} \right) {\bm S} \right)\\
&+ \sum_{k \in \cal K}\sum_{m\in {\cal M}_k}\left( \ell_{k,m}-\frac{\lambda_kN_0b_{k,m}t_{k,m}^{(1)}}{h_{k,m}}\left(2^{\frac{\ell_{k,m}}{t_{k,m}^{(1)}b_{k,m}}}-1\right) - \frac{\mu_{k,m}\xi_{k,m}C_{k,m}^3\ell_{k,m}^3}{\left(t_{k,m}^{(2)}\right)^2}\right)\\
&- \sum_{k \in \cal K}\sum_{m\in {\cal M}_k}\left( \frac{\mu_{k,m}N_0b_{k,m}t_{k,m}^{(3)}}{h_{k,m}}\left(2^{\frac{\beta\ell_{k,m}}{t_{k,m}^{(3)}b_{k,m}}}-1\right)+\rho_{k,m}\sum_{i=1}^3 t_{k,m}^{(i)}\right)\\
&+  \sum_{k \in \cal K} \left(\ell_{k,0}-\frac{\lambda_k\xi_{k,0}C_{k,0}^3\ell_{k,0}^3}{T^2}\right)+ \sum \limits_{n\in\cal N}\gamma_nP_{n} + \sum_{k \in \cal K}\sum_{m \in {\cal M}_k}\rho_{k,m} T.
\end{align*}
Accordingly, the dual function is given by
	\begin{align}\label{eq.func_dual}
	g(\{\lambda_{k}, \mu_{k,m}, \rho_{k,m}, \gamma_n\}) = \!\max_{\mbox{\tiny$\begin{array}{c}
\{\ell_{k,0}, \ell_{k,m}, t_{k,m}^{(i)}\}, \bm S\end{array}$}}& ~ {\cal L}\left(\bm S, \{\ell_{k,0}, \ell_{k,m}, t_{k,m}^{(i)}, \lambda_{k}, \mu_{k,m}, \rho_{k,m}, \gamma_n\}\right)\\
	{\text{s.t.}}~~~~~~&\eqref{eq.prob1_t}, ~\eqref{eq.prob1_l} ,~ {\text {and}}~\eqref{eq.prob1_S}.\notag
	\end{align}
%Note that the constraints $\bm \lambda>0$, $\bm \lambda>0$, and (\ref{eq.prob_dual}c) are to ensure the dual function $g(\bm \lambda, \bm \mu, \bm \rho, \bm \gamma )$ bounded from above (as proved in Appendix \ref{bounded_appendix}).
\begin{lemma}\label{lemma:1}
To ensure the dual function $g(\{\lambda_{k}, \mu_{k,m}, \rho_{k,m}, \gamma_n\})$ to be bounded from above, it must hold that $\lambda_{k}\!>\!0$, $ \mu_{k,m}\!>\!0$, and $\bm F\left(\!\{\lambda_{k},\mu_{k,m}, \gamma_n\right\}\!)\preceq \mv 0$, where $\bm F\left(\!\{\lambda_{k},\mu_{k,m}, \gamma_n\right\}\!)$ $\triangleq$ $ \sum_{k \in \cal K}(\lambda_kT\eta{\bm g}_{k,0}{\bm g}_{k,0}^{H}+\sum_{m \in {\cal M}_k}\mu_{k,m}T\eta{\bm g}_{k,m}{\bm g}_{k,m}^{H})-\sum_{n \in \cal N}\gamma_n{\bm \Sigma}_n $.
\end{lemma}

\begin{IEEEproof}
See Appendix~\ref{bounded_appendix}.
\end{IEEEproof}

Based on \eqref{eq.func_dual}, the dual problem of problem (P2.1) is expressed as
\begin{subequations}\label{eq.prob_dual}
	\begin{align}
\mathbf{(D2.1):}	\min_{\{\lambda_{k}, \mu_{k,m}, \rho_{k,m}, \gamma_n\}} ~&g(\{\lambda_{k}, \mu_{k,m}, \rho_{k,m}, \gamma_n\})\notag\\
	{\text{s.t.}}~~~~~~~~& \lambda_{k}\! >\!0,~ \!\mu_{k,m}\! >\! 0,~\! \rho_{k,m} \!\geq\! 0,~\!\gamma_{n}\! \geq \!0 ,~\!\forall m\in{\cal M}_k, ~\!k\in{\cal K},~\!n\in{\cal N}\label{multiples}\\
	&\bm F \left(\{\lambda_{k},\mu_{k,m}, \gamma_n\right\})\preceq \mv 0.\label{Fmultiples}
	\end{align}
\end{subequations}
As problem (P2.1) is convex and satisfies the Slater's condition, strong duality holds between problem (P2.1) and its dual problem (D2.1). In the following, we first evaluate the dual function $g(\{\lambda_{k}, \mu_{k,m}, \rho_{k,m}, \gamma_n\})$ under any given  $\{\lambda_{k}, \mu_{k,m}, \rho_{k,m}, \gamma_n\} \in {\cal X}$, and then find the optimal dual variables $\{\lambda_{k}, \mu_{k,m}, \rho_{k,m}, \gamma_n\}$ to minimize $g(\{\lambda_{k}, \mu_{k,m}, \rho_{k,m}, \gamma_n\})$, where the set $\cal X$ denotes the feasible set of $\{\lambda_{k}, \mu_{k,m}, \rho_{k,m}, \gamma_n\}$ specified by constraints \eqref{multiples} and \eqref{Fmultiples}. Let $\{\ell_{k,0}^{\rm opt1}, \ell_{k,m}^{\rm opt1}, t_{k,m}^{(i)\rm opt1}, \bm S^{\rm opt1}\}$ denote the optimal solution to problem (P2.1), and $\{\lambda_{k}^{\rm opt1}, \mu_{k,m}^{\rm opt1}, \rho_{k,m}^{\rm opt1}, \\ \gamma_{k,m}^{\rm opt1}\}$ denote the optimal dual solution to problem (D2.1).

\subsubsection{Derivation of Dual Function $g(\{\lambda_{k}, \mu_{k,m}, \rho_{k,m}, \gamma_n\})$}
 First, we obtain the dual function $g(\{\lambda_{k}, \mu_{k,m}, \rho_{k,m}, \gamma_n\})$ under any given  $\{\lambda_{k}, \mu_{k,m}, \rho_{k,m}, \gamma_n\} \in {\cal X}$ by solving problem  \eqref{eq.func_dual}. In this case, problem  \eqref{eq.func_dual} can be equivalently decomposed into the following $(K+M+1)$ subproblems
\begin{align}\label{eq.prob_sub1}
\max_{ \bm S\succeq 0}~&~ {\rm tr} \Big(\bm F\left(\{\lambda_{k},\mu_{k,m}, \gamma_n\right\}) \bm S\Big),
\end{align}
\begin{align}\label{eq.prob_sub2}
\max_{\ell_{k,0} \geq 0}~&~ \ell_{k,0}-\frac{\lambda_k\xi_{k,0}C_{k,0}^3\ell_{k,0}^3}{T^2},
\end{align}
%\begin{subequations}\label{eq.prob_sub3}
	\begin{align}
\max_{\{\ell_{k,m}, t_{k,m}^{(i)}\}} &~ ~ \ell_{k,m}-\frac{\lambda_kN_0b_{k,m}t_{k,m}^{(1)}}{h_{k,m}}\left(2^{\frac{\ell_{k,m}}{t_{k,m}^{(1)}b_{k,m}}}-1\right) - \frac{\mu_{k,m}\xi_{k,m}C_{k,m}^3\ell_{k,m}^3}{\left(t_{k,m}^{(2)}\right)^2}\notag\\ &-\frac{\mu_{k,m}N_0b_{k,m}t_{k,m}^{(3)}}{h_{k,m}}\left(2^{\frac{\beta\ell_{k,m}}{t_{k,m}^{(3)}b_{k,m}}}-1\right)
-\rho_{k,m}\sum_{i=1}^3 t_{k,m}^{(i)}\label{eq.prob_sub3}\\
{\text {s.t.}} ~~~&~~0 \leq \ell_{k,m},~0 \leq t_{k,m}^{(i)}\leq T,~\forall i \in \{1,2,3\}. \notag
	\end{align}
%\end{subequations}
Here, each subproblem in \eqref{eq.prob_sub2} and \eqref{eq.prob_sub3} corresponds to  user $k\in{\cal K}$ and helper $m \in {\cal M}_k$, respectively.
The optimal solutions to problems \eqref{eq.prob_sub1}--\eqref{eq.prob_sub3} denoted by $\bm S^*$, $\{\ell_{k,0}^*\}$, and $\{\ell_{k,m}\}$, respectively, are presented in the following respective lemmas.

\begin{lemma}\label{S_P_sub1}
Let ${\bm S}^*$ denote the optimal solution of problem \eqref{eq.prob_sub1}. To meet the optimality of problem \eqref{eq.prob_sub1}, $\bm S^*$ can be any positive semidefinite matrix that lies in the null space of $\bm F\left(\{\lambda_{k},\mu_{k,m}, \gamma_n\right\})\preceq \mv 0$. Thus, we choose $\bm S^{*} = \bm 0$ for evaluating the dual function only.\footnote{Notice that $\bm S^{*} = \bm 0$ is not feasible for the primal problem (P2.1) since the energy neutrality constraints in \eqref{energy consumption_users} and \eqref{energy consumption_helpers} are both violated. Therefore, an additional step is needed later in \eqref{prob.final} to obtain the feasible and thus optimal $\bm S^{{\rm opt1}}$.}
\end{lemma}

\begin{lemma}\label{Lk0_P_sub2}
Let $\{\ell_{k,0}^*\}$ denote the optimal solution to problem \eqref{eq.prob_sub2}, and it is given by
\begin{align}\label{lk0}
 \ell_{k,0}^* = \frac{T}{\sqrt{3\lambda_k\xi_{k,0}C_{k,0}^3}}, ~~\forall k \in \mathcal K.
\end{align}
\end{lemma}

\begin{IEEEproof}
Under any given $\lambda_k$, the objective function of problem \eqref{eq.prob_sub2} is a univariate function  with respect to $\ell_{k,0}$, $\forall \in\cal K$. Therefore, we obtain $\{\ell_{k,0}^*\}$ in \eqref{lk0} by checking its first-order derivative.
\end{IEEEproof}

\begin{lemma}\label{lem.t_l}
	The optimal solution to problem \eqref{eq.prob_sub3} is given by
     \begin{align*}
      \ell_{k,m}^*
      \begin{cases}
      =0,~&~{\rm if}~G\left(r^{(i)*}_{k,m},\lambda_k, \mu_{k,m},\rho_{k,m}\right)<0\\
      \in \left[0,\min_{i\in\{1,2,3\}} r^{(i)*}_{k,m}T\right],~&~{\rm if}~G\left(r^{(i)*}_{k,m},\lambda_k, \mu_{k,m},\rho_{k,m}\right)=0\\
      =\min_{i\in\{1,2,3\}}r^{(i)*}_{k,m}T, ~&~{\rm if}~ G\left(r^{(i)*}_{k,m},\lambda_k, \mu_{k,m},\rho_{k,m}\right)> 0,
      \end{cases}
    \end{align*}
	and
\begin{align}
t^{(i)*}_{k,m}=\frac{\ell^*_{k,m}}{r^{(i)*}_{k,m}}, \forall i\in\{1,2,3\},
\end{align}
     where
		\begin{align}
         &r^{(1)*}_{k,m}= \frac{b_{k,m}}{\ln2}\Big(1+W\Big(\frac{\rho_{k,m}h_{k,m}}{\lambda_kN_0b_{k,m}e}-\frac{1}{e}\Big)\Big),\notag\\
         &r^{(2)*}_{k,m}= \frac{1}{C_{k,m}}\Big(\frac{\rho_{k,m}}{2\mu_{k,m}\xi_{k,m}}\Big)^{\frac{1}{3}},\notag\\
         &r^{(3)*}_{k,m}= \frac{b_{k,m}}{\beta\ln2}\Big(1+W\Big(\frac{\rho_{k,m}h_{k,m}}{\mu_{k,m}N_0b_{k,m}e}-\frac{1}{e}\Big)\Big),\notag
		\end{align}
	and
	\begin{align}
	G\left(r^{(i)*}_{k,m},\lambda_k, \mu_{k,m},\rho_{k,m}\right) \triangleq &1-\frac{\lambda_kN_0b_{k,m}}{h_{k,m}r^{(1)*}_{k,m}}\left(2^{\frac{r^{(1)*}_{k,m}}{b_{k,m}}}-1\right)-\sum_{i=1}^3 \frac{\rho_{k,m}}{r^{(1)*}_{k,m}}-\frac{\lambda_kN_0b_{k,m}}{h_{k,m}r^{(3)*}_{k,m}}\left(2^{\frac{\beta r^{(3)*}_{k,m}}{b_{k,m}}}-1\right) \notag \\
	&-\mu_{k,m}{\xi_{k,m}C_{k,m}^3\left(r^{(1)*}_{k,m}\right)^2},~ \forall m \in {\cal M}_k,~ k\in{\cal K}, \label{G_r}
	\end{align}
with $e$ being Euler's number and $W(\cdot)$ being the Lambert $W$ function \cite{20}.
\end{lemma}
\begin{IEEEproof}
	See Appendix B.
\end{IEEEproof}
As stated in Lemma~\ref{lem.t_l}, if $G\left(r^{(i)*}_{k,m},\lambda_k, \mu_{k,m},\rho_{k,m}\right)=0$, the number of task input-bits $\ell_{k,m}^*\in\left[0,\min_{i\in\{1,2,3\}} r^{(i)*}_{k,m}T\right]$ for task offloading from user $k$ to helper $m \in{\cal M}_k $ is generally not a unique solution to problem \eqref{eq.prob_sub3}. In this case, we set $\ell_{k,m}^*=0$, $\forall m \in {\cal M}_k $, $k\in{\cal K}$, to facilitate the dual function evaluation. An additional procedure will be employed in the following to retrieve the optimal primal solution of $\{\ell_{k,0}^{\rm opt1}, \ell_{k,m}^{\rm opt1}\}$, together with $\bm S^{\rm opt1}$ and $t_{k,m}^{(i)\rm opt1}=\ell_{k,m}^{\rm opt1}/r^{(i)\rm opt1}_{k,m}$, $\forall m \in {\cal M}_k $, $k\in{\cal K}$.

By combining the obtained solutions in Lemmas \ref{S_P_sub1}--\ref{lem.t_l}, the dual function $g(\{\lambda_{k}, \mu_{k,m}, \rho_{k,m}, \gamma_n\})$ is finally evaluated under any given $\{\lambda_{k}, \mu_{k,m}, \rho_{k,m}, \gamma_n\}\in{\cal X}$.

\subsubsection{Obtaining  $\{\!\lambda_{k}^{\rm opt1}\!, \mu_{k,m}^{\rm opt1}, \rho_{k,m}^{\rm opt1}, \gamma_{k,m}^{\rm opt1}\!\}$  to Minimize Dual Function $g(\! \{\!\lambda_{k}\!, \mu_{k,m}, \rho_{k,m}, \gamma_{k,m}\!\}\!)$}
Next, we search over $\{\!\lambda_{k}, \mu_{k,m}, \rho_{k,m}, \gamma_n\!\} \!\in \!{\cal X}$ to minimize the dual function $g(\!\{\!\lambda_{k}, \mu_{k,m},
\rho_{k,m}, \gamma_{k,m}\!\}\!)$.
Generally, the dual function $g(\{\lambda_{k}, \mu_{k,m}, \rho_{k,m}, \gamma_n\})$ is convex but non-differentiable. As a result, the optimal dual solution $\{\lambda_k^{\rm opt1},\mu_{k,m}^{\rm opt1},\rho_{k,m}^{\rm opt1},\gamma_{k,m}^{\rm opt1}\}$ to problem (D2.1) can be obtained by subgradient-based methods such as the ellipsoid method \cite{19}.
To begin with, we choose a given $\{\lambda_{k}, \mu_{k,m}, \rho_{k,m}, \gamma_n\} \in {\cal X}$ as the center of the initial ellipsoid and set its volume to be sufficiently large to contain $\{\lambda_{k}^{\rm opt1}, \mu_{k,m}^{\rm opt1}, \rho_{k,m}^{\rm opt1}, \gamma_{k,m}^{\rm opt1}\}$.
Then, at each iteration, we update the dual variables $\{\lambda_{k}, \mu_{k,m}, \rho_{k,m}, \gamma_n\}$ based on the subgradients of both the objective and constraints functions, and accordingly establish a new ellipsoid with reduced volume. When the ellipsoid volume is below a certain threshold, the iteration terminates and the ellipsoid center is chosen to be the obtained  $\{\lambda_{k}^{\rm opt1}, \mu_{k,m}^{\rm opt1}, \rho_{k,m}^{\rm opt1}, \gamma_{k,m}^{\rm opt1}\}$.
To implement the ellipsoid method, it remains to determine the subgradients of both the objective and constraint functions. For the objective function in problem (D2.1), the subgradient with respect to $\{\lambda_{k}, \mu_{k,m}, \rho_{k,m}, \gamma_n\}$ is given as
\begin{align}\label{eq.subg}
&\Big[T\eta{\rm tr}( \mv S \bm g_{1,0}\bm g_{1,0}^H)  - \sum_{m\in {\cal M}_k}\frac{N_0b_{1,m}t_{1,m}^{(1)}}{h_{1,m}}\left(2^{\frac{\ell_{1,m}}{t_{1,m}^{(1)}b_{1,m}}}-1\right) - \frac{\xi_{1,0}C_{1,0}^3\ell_{1,0}^3}{T^2},\ldots,\notag \\
&T\eta{\rm tr}( \mv S \bm g_{K,0}\bm g_{K,0}^H)  - \sum_{m\in {\cal M}_K}\frac{N_0b_{K,m}t_{K,m}^{(1)}}{h_{K,m}}\left(2^{\frac{\ell_{K,m}}{t_{K,m}^{(1)}b_{K,m}}}-1\right) - \frac{\xi_{K,0}C_{K,0}^3\ell_{K,0}^3}{T^2},\ldots,\notag \\
&T \eta{\rm tr}( \mv S \bm g_{k,1}\bm g_{k,1}^H)- \frac{N_0b_{k,1}t_{k,1}^{(3)}}{h_{k,1}}\left(2^{\frac{\ell_{k,1}}{t_{k,1}^{(3)}b_{k,1}}}-1\right)
+\frac{\xi_{k,1}C_{k,1}^3\ell_{k,1}^3}{\left(t_{k,1}^{(2)}\right)^2},\ldots,\notag\\
&T \eta{\rm tr}( \mv S \bm g_{k,M}\bm g_{k,M}^H)- \frac{N_0b_{k,M}t_{k,M}^{(3)}}{h_{k,M}}\left(2^{\frac{\ell_{k,M}}{t_{k,M}^{(3)}b_{k,M}}}-1\right)
+\frac{\xi_{k,M}C_{k,M}^3\ell_{k,M}^3}{\left(t_{k,M}^{(2)}\right)^2},\ldots,\notag\\
&T-\sum_{i=1}^3 t_{1,m}^{(i)},\ldots,T-\sum_{i=1}^3 t_{k,M}^{(i)},P_{1}-{\rm tr}( \mv \Sigma_n \mv S ),\ldots,P_{N}-{\rm tr}( \mv \Sigma_n \mv S )\Big]^{\dagger},
\end{align}
The subgradients for the constraints in \eqref{multiples} are given by $\bm e_{k}$, $\forall k\in {\cal K}$, $\bm e_{K+m_1}$, $\forall m_1\in{\cal M}$, $\bm e_{K+M+m_2}$, $\forall m_2\in{\cal M}$, and $\bm e_{K+2M+n}$, $\forall n\in{\cal N}$, respectively, where $\bm e_l\in {\mathbb R}^{(K+2M+N)\times1}$ is the standard unit vector of all zero entries except for the $l$-th entry being one.
Regarding the constraint $\bm F\left(\{\lambda_{k},\mu_{k,m}, \gamma_n\right\})\preceq \mv 0$ in (\ref{eq.prob_dual}b), we establish the following lemma.

\begin{lemma}\label{lem.sub.F}
Let $\bm \nu \in \mathbb C^{N_tN\times1}$ denote the eigenvector corresponding to the smallest eigenvalue of $\bm F(\!\{\!\lambda_{k}, \mu_{k,m}, \gamma_n\!\}\!)$, i.e. $\bm \nu = \arg \min_{||\mv \varphi||=1}\mv\varphi^{H} \bm F(\!\{\!\lambda_{k}, \mu_{k,m}, \gamma_n\!\}\!)\mv\varphi$. The constraint $\bm F(\!\{\!\lambda_{k}, \mu_{k,m}, \gamma_n\!\}\!)\preceq \mv 0$ is equivalent to $\mv \nu^{H}\bm F\left(\{\lambda_{k},\mu_{k,m}, \gamma_n\right\})\mv \nu \geq 0$, and the subgradient of $\bm F\left(\{\lambda_{k},\mu_{k,m}, \gamma_n\right\})$ at given $\{\lambda_{k}, \mu_{k,m}, \rho_{k,m}, \gamma_n\}$ is
$\big[T\eta\mv \nu^{H}{\bm g}_{1,0}{\bm g}_{1,0}^{H}\mv \nu,\ldots,T\eta\mv \nu^{H}{\bm g}_{K,0}{\bm g}_{K,0}^{H}\mv \nu, T\eta\mv \nu^{H}\bm g_{k,1}{\bm g}_{k,1}^{H}\mv \nu, \ldots, T\eta\mv \nu^{H}\bm g_{k,M}{\bm g}_{k,M}^{H}\mv \nu,\\0, \ldots, 0, \mv \nu^{H} \bm \Sigma_1\mv \nu,\ldots,{\mv \nu^{H}} \bm \Sigma_N{\mv \nu} \big]^{\dagger}$.
\end{lemma}
\begin{IEEEproof}
	See Appendix C.
\end{IEEEproof}
Based on the subgradients in \eqref{eq.subg} and Lemma 5 and those for the constraints in \eqref{multiples}, the ellipsoid method can be applied to efficiently update $\{\lambda_{k}, \mu_{k,m}, \rho_{k,m}, \gamma_n\}$ towards $\{\lambda_{k}^{\rm opt1}, \mu_{k,m}^{\rm opt1}, \rho_{k,m}^{\rm opt1}, \\\gamma_{k,m}^{\rm opt1}\}$ for problem (D2.1).

\subsubsection{Optimal Solution to Problem {\rm (P2.1)}}\label{sec:P11_addition}
With the optimal dual solution  $\{\lambda_{k}^{\rm opt1}, \mu_{k,m}^{\rm opt1} , \rho_{k,m}^{\rm opt1},$\\
$ \gamma_{k,m}^{\rm opt1}\}$ at hand, it remains to determine the optimal solution to problem (P2.1). Specifically, by replacing $\{\lambda_{k}^*, \mu_{k,m}^*, \rho_{k,m}^*\}$ with $\{\lambda_{k}^{\rm opt1}, \mu_{k,m}^{\rm opt1}, \rho_{k,m}^{\rm opt1}\}$, we obtain the optimal $\{\ell_{k,0}^{\rm opt1}\}$ and $\{ r_{k,m}^{\rm opt1}\}$ based on Lemmas \ref{S_P_sub1} and \ref{lem.t_l}, respectively.
Due to the non-uniqueness of $\{\ell_{k,m}^*\}$ and $ \bm S^{*}$, one cannot obtain $\{\ell_{k,m}^{\rm opt1}\}$ and $ \bm S^{\rm opt1}$ directly.
By substituting $\ell_{k,0}^{\rm opt1}$, $t_{k,m}^{(1)\rm opt1} = \frac{\ell_{k,m}^{\rm opt1}}{r_{k,m}^{(1)\rm opt1}}, t_{k,m}^{(2)\rm opt1} = \frac{\ell_{k,m}^{\rm opt1}}{r_{k,m}^{(2)\rm opt1}}$, and $t_{k,m}^{(3)\rm opt1} = \frac{\ell_{k,m}^{\rm opt1}}{r_{k,m}^{(3)\rm opt1}}, ~\forall m\in {\cal M}_k,~k\in{\cal K}$, in problem (P2.1), we solve the following problem to obtain $\bm S^{\rm opt1}$ and $\{\ell_{k,m}^{\rm opt1}\}$:
\begin{align} \label{prob.final}
\max_{ \bm S \succeq \bm 0, \{\ell_{k,m}\}} ~&\sum_{k \in \cal K}\sum_{m\in {\cal M}_k} \ell_{k,m} \\
{\rm s.t.}~~~~~ &E_{k,0}^{\rm tx} + \frac{T}{\sqrt{27\xi_{k,0}\left(\lambda_k^{\rm opt} C_{k,0}\right)^{3}}}\leq T \eta {\rm tr}(\bm g^H_{k,0} \bm S\bm g_{k,0}),~\forall k\in{\cal K}  \notag\\
& E_{k,m}^{\rm tx}+\xi_{k,m}C_{k,m}^3\ell_{k,m}r_{k,m}^{(2)\rm opt1}\leq T \eta {\rm tr}(\bm g_{k,m}^H \bm S\bm g_{k,m}),~\forall m\in {\cal M}_k,~k\in{\cal K}\notag\\
&{\rm tr}( \mv \Sigma_n \bm S  ) \leq P_{n}, ~\forall n \in \cal N \notag\\
& \sum_{m \in {\cal M}_k}\frac{\ell_{{k,m}}}{r_{k,m}^{(i)\rm opt1}} \leq T,~ \forall i\in\{1,2,3\},~k \in {\cal K} \notag\\
&\ell_{k,m}\geq 0,~\forall i\in\{1,2,3\},~\forall m\in {\cal M}_k, ~\forall k\in{\cal K} \notag.
\end{align}
Note that the problem \eqref{prob.final} is a semi-definite program (SDP) that can be efficiently solved by standard convex optimization tools such as CVX \cite{19}. With $\{\ell_{k,m}^{\rm opt1}\}$ obtained, we accordingly have $t_{k,m}^{(i){\rm opt1}}=\ell_{k,m}^{\rm opt1}/r^{(i) \rm opt1}_{k,m}$, $\forall i\in\{1,2,3\}$, $m\in{\cal M}_k$, $k \in {\cal K}$.
Then, by combining $\{\ell_{k,m}^{\rm opt1}, t_{k,m}^{(i)\rm opt1}, \bm S^{\rm opt1}\}$, together with $\{\ell_{k,0}^{\rm opt1}\}$, the optimal solution to (P2.1) is finally found. To summarize, we present Algorithm 1 to optimally solve problem (P2.1).
\begin{algorithm}
	\caption{$\textit{: Proposed Solution for Optimally Solving Problem}$ (P2.1)}
	\begin{algorithmic}[1]
		\item[1:] {\bf Initialization}: Given an ellipsoid ${\cal E}((\{\lambda_{k}, \mu_{k,m}, \rho_{k,m}, \gamma_n\}), \bm A)$ containing $ \{\lambda_k^{\rm opt1}, \mu_{k,m}^{\rm opt1}, \rho_{k,m}^{\rm opt1}, \gamma_{n}^{\rm opt1}\}$,  where $\{\lambda_{k}, \mu_{k,m}, \rho_{k,m}, \gamma_n\}$ is the center of ${\cal E}$ and the positive definite matrix ${\mv A}$ characterizes the size of ${\cal E}$.
		\item[2:]{\bf Repeat}:
		\begin{enumerate}
			\item Obtain $\{\ell_{k,m}^{*}, \ell_{k,0}^*, t_{k,m}^{(1)*}, t_{k,m}^{(2)*}, t_{k,m}^{(3)*}, \bm S^{*}\}$ under $\{\lambda_k, \mu_{k,m}\, \rho_{k,m}, \gamma_{n}\}$ based on Lemmas \ref{S_P_sub1}--\ref{lem.t_l}.
			\item Compute the subgradients of $ g(\{\lambda_{k}, \mu_{k,m}, \rho_{k,m}, \gamma_n\})$ based on \eqref{eq.subg} and those of \eqref{multiples} and \eqref{Fmultiples} based on Lemma 5, and update $\{\lambda_k, \mu_{k,m}, \rho_{k,m}, \gamma_{n}\}$ by using the ellipsoid method.
		\end{enumerate}
		\item[3:]{\bf Until:} {$\{\lambda_k, \mu_{k,m}, \rho_{k,m}, \gamma_{n}\}$ converges within the desired accuracy.}	
		\item[4:] Set $\{\lambda_k^{\rm opt1}\} \leftarrow \{\lambda_k\}$, $\{\mu_{k,m}^{\rm opt1}\} \leftarrow \{\mu_{k,m}\}$, $\{\rho_{k,m}^{\rm opt1}\} \leftarrow \{\rho_{k,m}\}$, and $\{\gamma^{\rm opt1}_n\} \leftarrow \{\gamma_n\}$, and obtain the optimal \{$\ell_{k,0}^{\rm opt1}$\} and $ \{r_{k,m}^{(1)\rm opt1}, r_{k,m}^{(2)\rm opt1}, r_{k,m}^{(3)\rm opt1}\}$.
        \item[5:] Obtain $\{\ell_{k,m}^{\rm opt1}, t_{k,m}^{(1){\rm opt1}}, t_{k,m}^{(2){\rm opt1}}, t_{k,m}^{(3){\rm opt1}}, \bm S^{\rm opt1}\}$ by solving problem \eqref {prob.final} with $ \{r_{k,m}^{(1)\rm opt1}, r_{k,m}^{(2)\rm opt1}, r_{k,m}^{(3)\rm opt1}\}$.
		\item[6:] {\bf Output}: The optimal solution of $\{\ell_{k,m}^{\rm opt1}, \ell_{k,0}^{\rm opt1}, t_{k,m}^{(1)\rm opt1}, t_{k,m}^{(2)\rm opt1}, t_{k,m}^{(3)\rm opt1}, \bm S^{\rm opt1}\}$ to problem (P2.1).
	\end{algorithmic}
\end{algorithm}
\subsection{Optimization over  $ \{\ell_{k,0}, \ell_{k,m}, b_{k,m}, \bm S\}$ under Given $\{t_{k,m}^{(i)}\}$}
In this subsection, we consider problem (P2) under given time allocation $\{t_{k,m}^{(i)}\}$, which is reduced to
%\begin{subequations}\label{eq.prob2.2}
	\begin{align*}
\mathbf{(P2.2):}~\psi_2(\{{\cal{M}}_k, t_{k,m}^{(i)}\}) = \max_{\{\ell_{k,0}, \ell_{k,m}, b_{k,m}\},\bm S}~&  \sum_{k \in \cal K}\left(\ell_{k,0}+\sum_{m\in {\cal M}_k} \ell_{k,m}\right)\\
{\rm s.t.}~~~~~~~ &\eqref{power}, ~ \eqref{frequence},~\eqref{duration} ,~\eqref{energy consumption_users}, ~\eqref{energy consumption_helpers}, ~{\text{and}} ~\eqref{eq.prob1_l}-\eqref{eq.prob1_S}.
	\end{align*}
%\end{subequations}
Note that problem (P2.2) is a convex optimization problem that can be solved optimally by leveraging the Lagrange duality method.
The procedure for solving problem (P2.2) is similar as that for problem (P2.1).
We denote the Lagrange multipliers as $\bar{a}_k\geq 0$, $\forall k \in \cal K$, $\bar{c}_{k,m}\geq 0$, $\bar{d}\geq 0$ and $\bar{j}_{k,m}\geq 0$, $\forall m \in {\cal M}_k$, $k \in \cal K$, for \eqref{power}, \eqref{frequence}, \eqref{energy consumption_users}, and \eqref{energy consumption_helpers}.
The optimal solution to problem (P2.2) is immediately presented in the following proposition and we omit the details for brevity. Let $ \{\ell_{k,0}^{opt2}, \ell_{k,m}^{opt2}, b_{k,m}^{opt2}, \bm S^{opt2}\}$ denote the optimal solution to problem (P2.2).

\begin{proposition}\label{lem.b}
The optimal solution of $\{\ell_{k,0}^{\rm opt2}\}$ to problem (P2.2) is given by
\begin{align}\label{eq.opt_l02}
\ell_{k,0}^{\rm opt2} = \frac{T}{\sqrt{\bar{a}_{k}^{\rm opt2}\xi_{k,m}C_{k,m}^3}}, ~\forall m \in {\mathcal M}_k,~ k \in \mathcal K.
\end{align}
The optimal solution of $\{\ell_{k,m}^{\rm opt2}, b_{k,m}^{\rm opt2}\}$ is obtained by solving the following equations
%\begin{subequations}\label{eq.sub_kkt2}
\begin{align}
	&1-\frac{\bar{a}_k^{\rm opt2}N_0\ln 2}{h_{k,m}}2^{\frac{\bar E^{\rm opt2}_{k,m}}{N_0t_{k,m}^{(1)}}}-\frac{\beta \bar{c}_{k,m}^{\rm opt2}N_0\ln 2}{ h_{k,m}}2^{\frac{\beta \bar E^{\rm opt2}_{k,m}}{N_0t_{k,m}^{(3)}}}-{\frac{3\bar{c}_{k,m}^{\rm opt2}\xi_{k,m}C_{k,m}^3 \left(\ell_{k,m}^{{\rm opt2}}\right)^2}{t_{k,m}^{(2)}}}= 0, \notag\\
	& \frac{\bar{a}_k^{\rm opt2}N_0t_{k,m}^{(1)}}{h_{k,m}}2^{\frac{\bar E_{k,m}^{\rm opt2}}{t_{k,m}^{(1)}}}\left(\frac{\ln2\bar E_{k,m}^{\rm opt2}}{t_{k,m}^{(1)}} -1\right)+\frac{\bar{a}_k^{\rm opt2}N_0t_{k,m}^{(3)}}{h_{k,m}}+
\frac{\bar{c}_{k,m}^{\rm opt2}N_0t_{k,m}^{(3)}}{h_{k,m}} 2^{\frac{\beta \bar E_{k,m}^{\rm opt2}}{t_{k,m}^{(3)}}}\left(\frac{\beta \ln2\bar E^{\rm opt2}_{k,m}}{t_{k,m}^{(3)}}-1\right)\notag\\
&+\frac{\bar{c}_{k,m}^{\rm opt2}N_0t_{k,m}^{(3)}}{h_{k,m}} -\bar d^{\rm opt2}  =  0,\notag
\end{align}
%\end{subequations}
where $\bar E^{\rm opt2}_{k,m} \triangleq \frac{\ell_{k,m}^{\rm opt2}}{b_{k,m}^{\rm opt2}}$, $\forall m\in{\cal M}_k, k\in{\cal K}$. Here, we let $\bar{j}_{n}^{\rm opt2}\geq 0$, $\bar{d}^{\rm opt2}\geq 0$, $\bar{a}_{k}^{\rm opt2}\geq 0$,  and $\bar{c}_{k,m}^{\rm opt2}\geq 0$, $\forall m \in {\cal M}_k$, $k \in \cal K$ denote the optimal Langrange multipliers associated with constraints in \eqref{power}, \eqref{frequence}, \eqref{energy consumption_users}, and \eqref{energy consumption_helpers}, respectively, which can be obtained by solving the dual problem of problem (P2.2) via the ellipsoid method.
The optimal energy transmit covariance matrix $ \bm S^{\rm opt2}$ to problem (P2.2) is obtained by solving an SDP by substituting $\{ \ell_{k,0}^{\rm opt2}, \ell_{k,m}^{\rm opt2}, b_{k,m}^{\rm opt2}\}$ back into problem (P2.2), similarly as that for problem \eqref{prob.final}.

\end{proposition}

\subsection{Alternating Optimization}
We now present the  alternating-optimization-based approach to solve problem (P2) by iteratively solving problems (P2.1) and (P2.2) in an alternating manner. In each iteration, we first solve problem (P2.1) under given $\{b_{k,m}\}$ to update $\{t_{k,m}^{(i)}, \ell_{k,m}, \ell_{k,0}, \bm S\}$, and then solve problem (P2.2) under given $\{t_{k,m}^{(i)}\}$ to update $\{b_{k,m}, \ell_{k,m}, \ell_{k,0}, \bm S\}$.
For initialization, we consider an identical frequency bandwidth $b_{k,m} = \frac{B}{\sum_{k=1}^K |{\mathcal M}_k|}$ for computation offloading between each user-helper pair.
Since problems (P2.1) and (P2.2) are convex and can be optimally solved, the objective value of problem (P2) is monotonically nondecreasing after each iteration.
Therefore, the proposed alternating-optimization-based algorithm converges towards a stationary sub-optimal solution to problem (P2).
%\begin{algorithm}
%	\caption{$\textit{for Solving Problem}$ (P2)} %?????
%	\begin{algorithmic}[1]
%		\State {\bf Initialization}: Given the $\bm b^{(0)} \triangleq \{b_{k,m}^{(0)}\}$, and $b_{k,m}^{(0)} = \frac{B}{M}$, $\forall m \in {\cal M}_k$, $k \in \cal K$. Let $i = 0$.
%		\State {\bf Repeat}:
%		\begin{enumerate}
%			\item Solve (P2.1) to obtain ($ \bm S^{(i)*}$, $\{\bm t_{k,m}^{(i)*}\}$, $\{\ell_{k,m}^{(i)*}, \ell_{k,0}^{(i)*}\}$) with initial $\{b_{k,m}^{(i)*}\}$.
%			\item Solve (P2.2) to obtain ($ \bm S^{(i+1)*}$, $\{\bm t_{k,m}^{(i+1)*}\}$, $\{\ell_{k,m}^{(i+1)*}, \ell_{k,0}^{(i+1)*}\}$) with above results of $\{\bm t_{k,m}^{(i)*}\}$.
%            \item Update $i=i+1$ ;
%		\end{enumerate}
%		\State {\bf Until} the growth of the objective value $\psi(\{{\cal{M}}_k, b_{k,m}\})$-$\psi(\{{\cal{M}}_k, \bm t_{k,m}\})$ satisfy the accuracy $\varepsilon$, which is the $\frac{\max({\psi(\{{\cal{M}}_k, b_{k,m}\}), \psi(\{{\cal{M}}_k, \bm t_{k,m}\})})}{1000}$.
%		\State {\bf Output}: With the $\{\ell_{k,m}^{\rm opt}, \ell_{k,0}^{\rm opt}\}, \{\bm t_{k,m}^{\rm opt}\}, \{b_{k,m}
%^{\rm opt}\}$ and $ \bm S^{\rm opt}$, the optimal solution is found.
%	\end{algorithmic}
%\end{algorithm}

\section{User-Helper Pairing Schemes}\label{sec:pairing}
In practice, the pairing relationship between the users and helpers are needed to be established and adjusted subsequently due to the variation of the related wireless communication channels and/or computation tasks. In this section, we present two user-helper pairing schemes by using the exhaustive search and greedy selection, respectively, and also present a benchmark scheme based on the channel power gains between the users and helpers only.

\subsection{Optimal User-Helper Pairing via Exhaustive Search}
In the optimal exhaustive search-based pairing scheme, we first enumerate all the user-helper pairing sets between the $K$ users and $M$ helpers, and then choose the optimal user-helper pairing sets $\{{\cal{M}}_k^{\rm opt}\}$ as the one achieving the maximum sum computation rate for  problem (P1). Hence, the optimal $\{{\cal{M}}_k^{\rm opt}\}$ is obtained as
\begin{align}\label{exhaustive}
\{{\cal{M}}_k^{\rm opt}\} = \arg \max_{\{{\cal M}_k\}} ~\psi\left(\{{\cal{M}}_k\}\right).
\end{align}
Recall that $\psi(\{{\cal M}_k\})$ (given in problem (P2)) is the obtained sum computation rate under given $\{{\cal M}_k\}$. For the exhaustive search problem \eqref{exhaustive}, there are a total of $K^M$ user-help pairing candidates. Accordingly, one has to solve problem (P2) for $K^M$ times each with given user-helper pairing set $\{{\cal M}_k\}$.
The computational complexity of the exhaustive search scheme exponentially increases with respect to the number of users and helpers, and thus is not acceptable for practical implementation.
Therefore, it is desirable to develop low-complexity user-helper pairing algorithms with slight compromise in the sum computation rate performance.

\subsection{Greedy-Based User-Helper Pairing}
In this subsection, to reduce the complexity, we develop a greedy-based scheme to obtain the user-helper pairing sets efficiently. This scheme is implemented in an iterative manner as follows. To start with, we consider the case without user-helper pairing, such that all the users execute their computation tasks locally by themselves, which can be expressed as
\begin{align}
{\cal M}_k = \phi, ~\forall k \in {\cal K}.
\end{align}
Next, it takes $M$ iterations to determine the paired user for each helper.
In each iteration $i\in \{1,..., M\}$, we choose one helper $m \in {\cal M} \setminus (\cup_{k\in \cal K} {\cal M}_k)$ from the set of non-pairing helpers, and pair it with any one user $k$.
Correspondingly, we update $\overline {\cal M}_{k(m)} \gets {\cal M}_k \cup \{m\}$ and keep other sets $\{{\cal M}_j\}_{j\neq k}$ unchanged. In this case, we solve problem (P2) under given $\{{\cal M}_1, \ldots, {\overline {\cal M}_{k(m)}}, \ldots, {\cal M}_K\}$ to obtain the corresponding sum computation rate as $\psi( \{{\cal M}_1, \ldots, \\\overline {\cal M}_{k(m)}, \ldots, {\cal M}_K \})$.
Then we calculate the sum computation rate under these updated user-helper pairing sets, and obtain $\{{\cal M}_k\}$ as the one achieving the maximal sum computation rate. In other words, we have
\begin{align}
(m^{(i)}, k^{(i)}) = \arg \max_{m\in{\cal M}_k, k\in{\cal K}} \psi(\{ {\cal M}_1, \ldots, \overline {\cal M}_{k(m)}, \ldots, {\cal M}_K\} ),
\end{align}
and we update the user-helper pairing sets as $\{{\cal M}_k\} \gets \{{\cal M}_1, \ldots, {\overline {\cal M}_{k^{(i)}(m^{(i)})}}, ..., {\cal M}_K\}.$

Note that adding one additional helper in one certain iteration may not be able to further increase the sum computation rate of the users. In this case, the proposed greedy-based algorithm terminates immediately.
At each iteration $i$, problem (P2) needs to be solved for at most $K (M-i)$ times. Therefore, problem (P2) needs to be solved for $\frac{KM(M-1)}{2}$ times in the worst case.  As compared to the optimal exhaustive search scheme, the computational complexity of the proposed greedy-based scheme is significantly reduced, especially when the number of users is large. As a summary, we present the proposed greedy-based user-helper pairing scheme as Algorithm \ref{greedy}.

\begin{algorithm}
	\caption{\textit{: Proposed Greedy-Based User-Helper Pairing}}
\begin{algorithmic}[1]
		\item[1:] {\bf Initialization}: Initialize ${\cal M}_k$ = $\phi$ as the empty helper set chosen by user $k$, $ ~\forall k \in \cal K$. Let $i$ be the iteration index as $i=0$, and $\psi^{(i*)}$ be the users' sum computation rate by their local computing. Let $i=1$.
       \item[2:] {\bf Repeat}:
        \begin{enumerate}
        \item $k=1$.
		\item {\bf Repeat}:
		\begin{enumerate}
            \item $m=1$.
			\item$\textbf{Repeat}$:
                 \begin{enumerate}
            \item $\overline {\cal M}_{k(m)} \gets {\cal M}_k \cup \{m\}$;
            \item Solve problem (P2) under given $\{{\cal M}_1, \ldots, {\overline {\cal M}_{k(m)}}, \ldots, {\cal M}_K\}$ to obtain $\psi^{(i)}(\{ {\cal M}_1, \ldots, \overline {\cal M}_{k(m)}, \ldots, {\cal M}_K \})$.
            \item $k = k+1$.
                  \end{enumerate}
            \item$\textbf{Until}$: $k = K$.
		    \item $m = m+1$.
		\end{enumerate}
		\item  {\bf Until} $m = |{\cal M} \setminus (\cup_{k\in \cal K} {\cal M}_k)|$.
        \item  Obtain $\psi^{(i*)} = \max_{m \in {\cal M}_k, k \in {\cal K}} \psi^{(i)}( \{{\cal M}_1, \ldots, \overline {\cal M}_{k(m)}, \ldots, {\cal M}_K \})$ and $(m^{(i)}, k^{(i)}) = \arg \max_{m\in{\cal M}_k, k\in{\cal K}} \psi(\{ {\cal M}_1, \ldots, \overline {\cal M}_{k(m)}, \ldots, {\cal M}_K\} )$, and then update $\{{\cal M}_k\} \gets \{{\cal M}_1, \ldots, {\overline {\cal M}_{k^{(i)}(m^{(i)})}}, \ldots, {\cal M}_K\}$.
        \item $i = i+1$.
       \end{enumerate}
        \item[3:] {\bf Until} $i = M$.
        \item[4:] Obtain $\{{\cal M}_k^{\rm opt}\}$.
	\end{algorithmic}\label{greedy}
\end{algorithm}

%\begin{algorithm}
%	\caption{channel based algorithm for pairing} %?????
%	\begin{algorithmic}[1]
%		\State obtain all $M \times K$ channels condition
%        \State {\bf Initialize} $m = 0$,
%        \State {\bf repeat}
%        \begin{enumerate}
%           \item $ m= m+1$;
%		   \item Compare $K$ channels state between helper $m$ with users
%       \end{enumerate}
%        \State {\bf Until} $m = M$.
%        \State All the strategies are obtain the optimal ${\cal M}_k$, $\forall k \in \cal K$, the optimal selection strategy is found.
%	\end{algorithmic}
%\end{algorithm}

\subsection{Channel-Based User-Helper Pairing}
In this subsection, we present the user-helper pairing scheme based on the channel power gains between the users and helpers for benchmarking purposes. In particular, we consider that each helper is always paired with the user with the largest channel power gain. For each helper $m\in \cal M$, we set it to be paired with user $\bar{k}$, if ${\bar{k}} = \arg\max_{k\in\cal K} h_{k,m}$.
This channel-based user-helper pairing scheme has a considerably low computational complexity but compromised performance, as will be shown in the simulation results next.
%$$ f(x)=\left\{
%\begin{aligned}
%&m \in {\cal M}_k \quad \quad \quad{\tt if} \quad k = \argmax\{h_{k,m}\} \\
%&m \not\in {\cal M}_k \quad \quad \quad \tt otherwise
%\end{aligned}
%\right.
%$$

\section{Numerical Results}\label{sec:results}
In this section, we present numerical results to evaluate the performance of the proposed wireless powered D2D cooperative computing designs. In the context of wireless powered D2D-enabled cooperative computing, there are $N=2$ ETs with $N_t=4$ antennas. For all the $K$ users and $M$ helpers, we set the EH efficiency coefficient as $\eta = 0.8$, the switch capacitance $\xi_{k,0} = 10^{-28}$, the required CPU cycles per task input-bits $C_{k,0}= C_{k,m}=10^3$, $\forall m \in {\mathcal M}_k,~ k \in \cal K$. The PSD of AWGN at receivers is $N_0 =10^{-15}$ and the system bandwidth for offloading is $B=3$ MHz.
All the wireless channels between the ETs and users, users and helpers are modeled as independent and identically distributed Rayleigh fading with an average power loss of ${\rm PL}_{0}\times d^{-\theta}$, where ${\rm PL}_{0}= -30$ dB is the channel power loss at a reference distance of 1 meter (m), $d$ denotes the distance from the transmitter to receiver $i$, and $\theta=3$ is the path-loss exponent.
In addition, we include following two benchmark schemes for performance comparison.

\begin{enumerate}
\item {\it Optimal beamforming with local computing}: The users complete their computation tasks by local computing only. This scheme corresponds to solving problem (P2) by setting $\ell_{k,m} = 0$ and $t_{k,m}^{(1)} = t_{k,m}^{(2)} = t_{k,m}^{(3)} = 0$, $\forall m \in{\cal M}_k$, $k \in \cal K$, which can be formulated as
\begin{align}\label{eq.user}
\max_{\{\ell_{k,0}\geq0\},\bm S }~&  \sum_{k \in \cal K} \ell_{k,0}\\
\quad \quad {\rm s.t.}~~~& \frac{\xi_{k,0}C_{k,0}^3\ell_{k,0}^3}{T^2}\leq T \eta{\rm tr}( \bm S {\bm g}_{k,0}{\bm g}_{k,0}^{H}),~\forall k\in{\cal K}\notag\\
&  {\rm tr}( \mv \Sigma_n \bm S  ) \leq P_{n}, \bm S \succeq \bm 0,~\forall n\in {\cal N}.\notag
\end{align}

\item {\it Uniform beamforming with optimized offloading}: In this scheme, we set $ \bm {\Sigma_n} \bm S= \frac{P_{n}}{N_t}\boldsymbol{I}, ~\forall n\in {\cal N}$, and the joint communication and computation resource allocation for D2D-enabled cooperative computing is formulated as
\begin{align}\label{eq.prob1_US}
\max_{
 \{{\cal M}_k, \ell_{k,0}, \ell_{k,m}, b_{k,m}, t_{k,m}^{(i)}\}}~&  \sum_{k \in \cal K} \left(\ell_{k,0}+\sum_{m\in {\cal M}_k} \ell_{k,m}\right)\\
{\text {s.t.}}~~~~~~~~~~~&\eqref{frequence}, ~\eqref{duration} ,~\eqref{energy consumption_users}, ~\eqref{energy consumption_helpers},~{\text {and}} ~ \eqref{eq.prob_M}-\eqref{eq.prob1_S}.\notag
\end{align}
\end{enumerate}
Both problems \eqref{eq.user} and \eqref{eq.prob1_US} can be solved similarly as those in Section \ref{sec:design}.

\begin{figure}
\begin{minipage}[t]{0.5\linewidth}
 \includegraphics[width=3.3in]{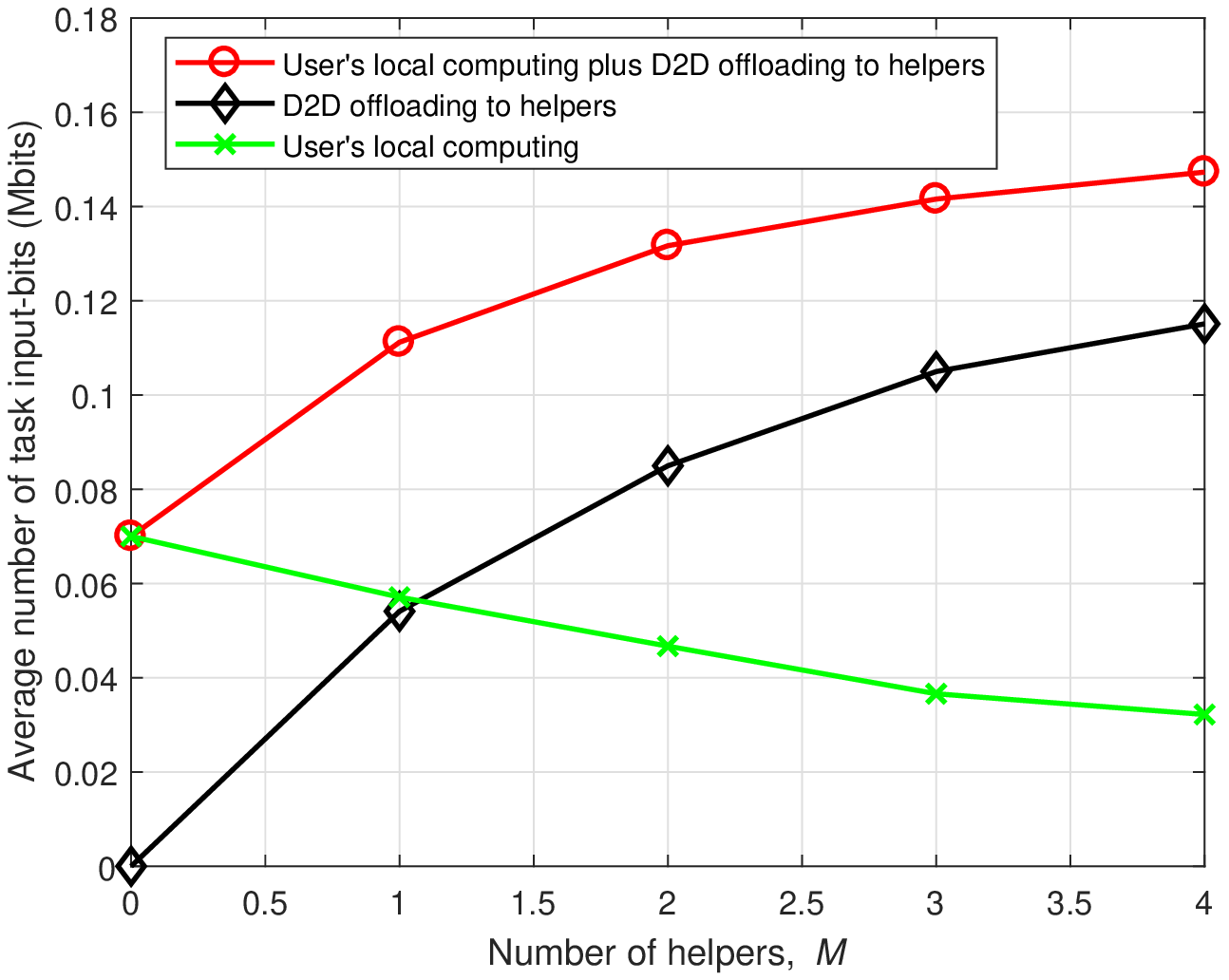}
\caption{The average number of task input-bits versus the \protect\\number of helpers $M$, where $K$=1 and $N$=2.}\label{fig:single_cap}
\end{minipage}%
\begin{minipage}[t]{0.5\linewidth}
 \includegraphics[width=3.3in]{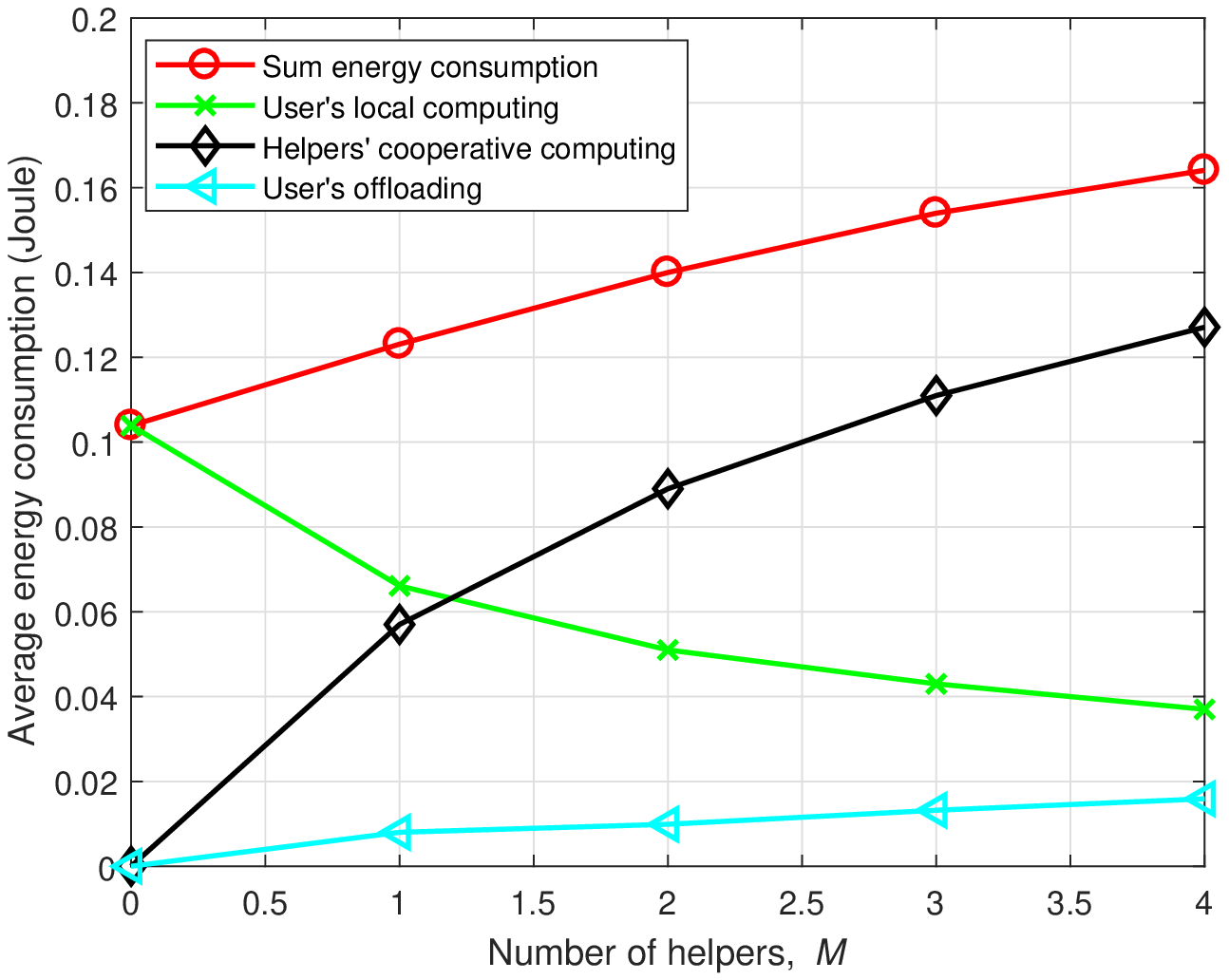}
 \caption{The average energy consumption versus the number of \protect\\helpers $M$, where $K$=1 and $N$=2.}\label{fig:single_ener}
\end{minipage}
\end{figure}

First, we consider a special single-user case with $K=1$. Figs. \ref{fig:single_cap} and \ref{fig:single_ener} show the user's average number of computation task input-bits and the corresponding energy consumption versus the number of helpers $M$, where $N=2$, the block duration $T = 0.3$ second (s)  and the power budget $P=6$ Watt (W). In Fig.~\ref{fig:single_cap}, it is observed that the number of task input-bits computed by the helpers increases as $M$ increases, but it is not true for the user's local computing. This is because that the two ETs adjust their beamforming designs with the objective of maximizing the sum computation task input-bits executed by the user and the increasing helpers, such that the user prefers D2D offloading to local computing. As expected, it is observed in Fig. \ref{fig:single_ener}, the user spends increasing energy amount in D2D offloading to the helpers and decreasing energy amount in local computing as $M$ grows.  It is also observed in Fig. \ref{fig:single_ener} that the helpers' energy consumption in executing the user's offloaded tasks increases as $M$ increases.

\begin{figure}
\begin{minipage}[t]{0.5\linewidth}
 \includegraphics[width=3.3in]{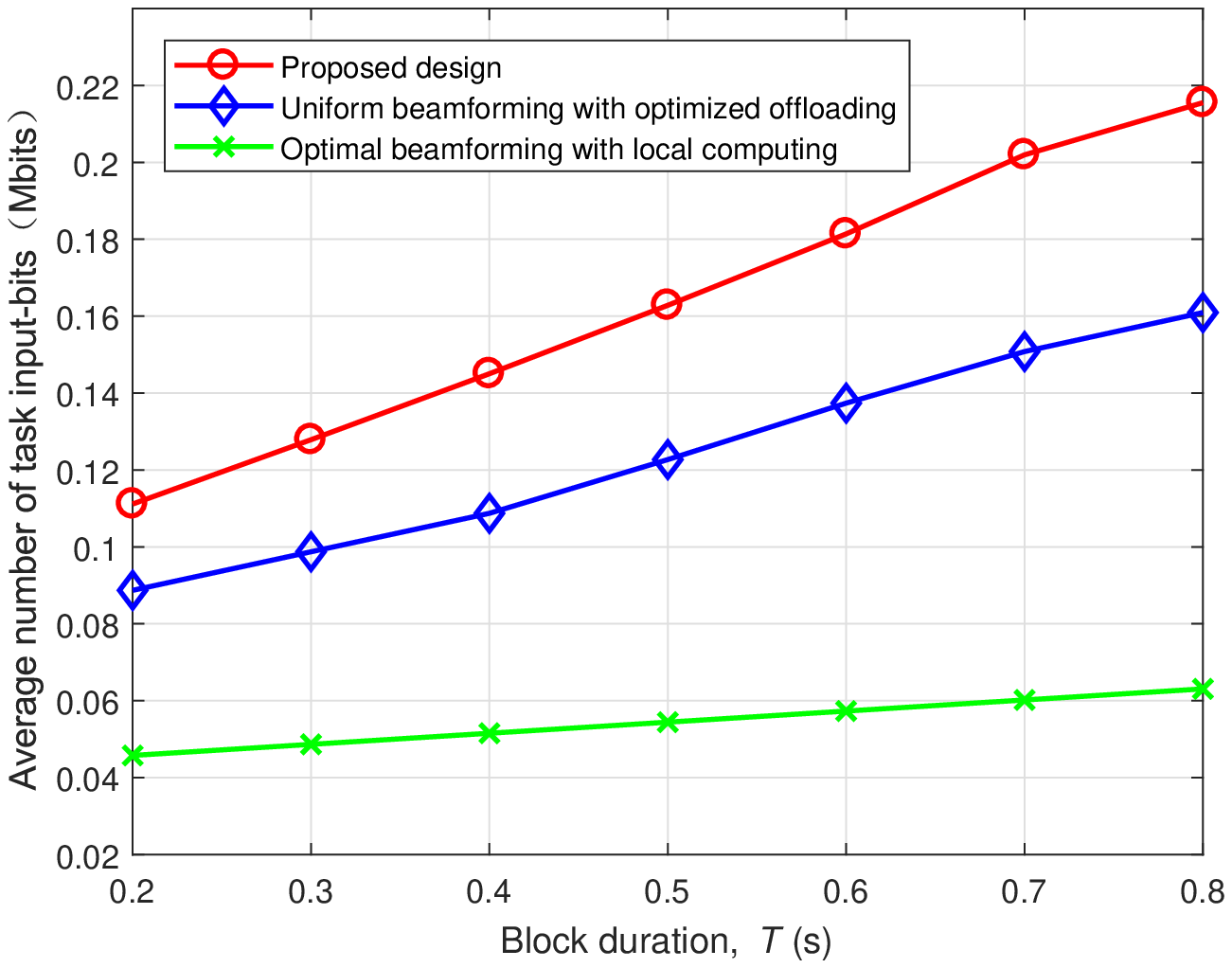}
 \caption{ The average number of task input-bits versus the block \protect\\duration $T$, where $K$=1, $M$=3, and $N$=2.}\label{fig:user3_result_t}
\end{minipage}
\begin{minipage}[t]{0.5\linewidth}
 \includegraphics[width=3.3in]{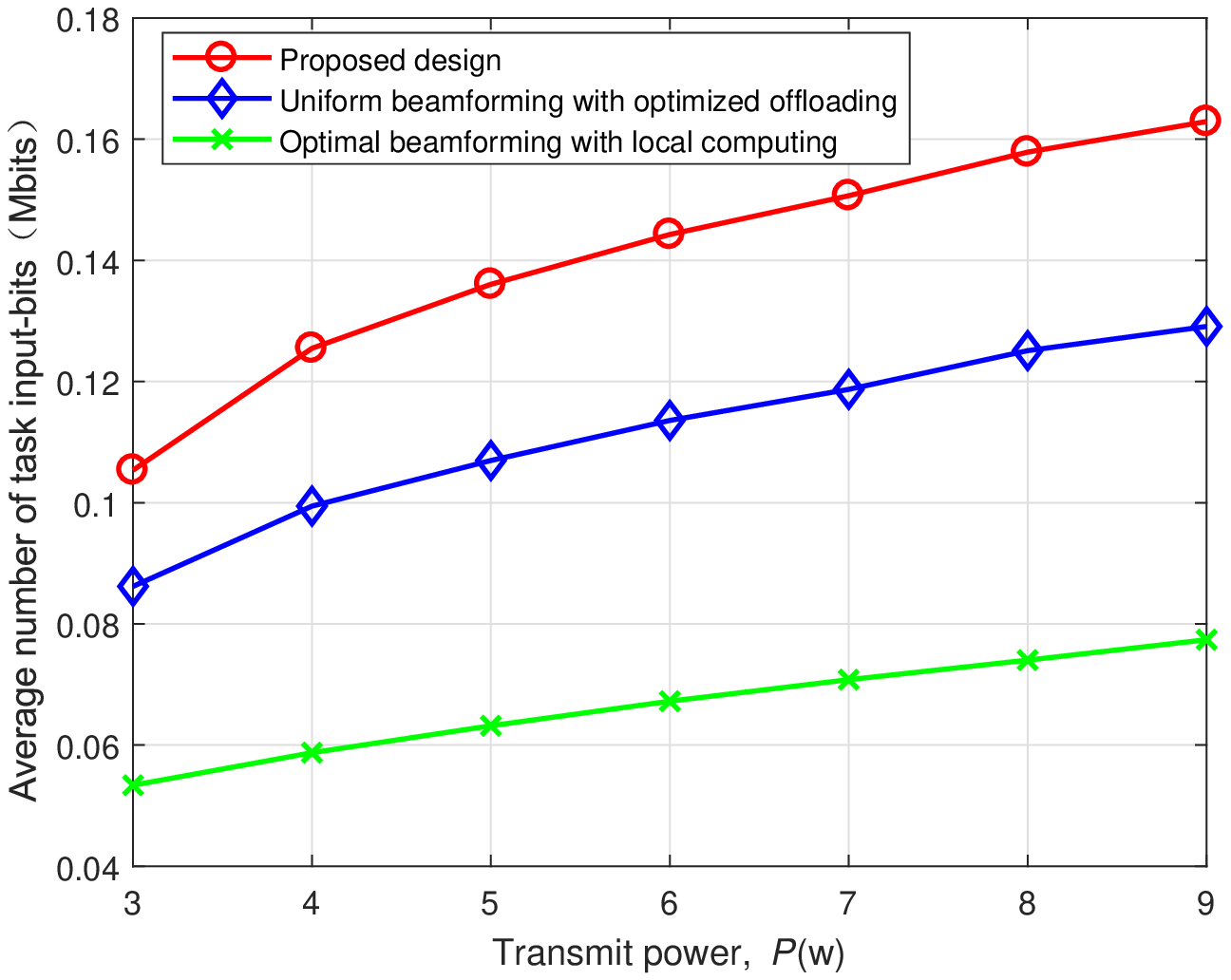}
 \caption{ The average number of task input-bits versus the ET's \protect\\ transmit power budget $P$, where $K$=1, $M$=3, and $N$=2.}\label{fig:user3_result_p}
\end{minipage}%
\end{figure}

Furthermore, Figs.~\ref{fig:user3_result_t} and \ref{fig:user3_result_p} show the average number of computation task input-bits versus the block duration $T$ and the power budget $P$, respectively, where $M=3$ and $N=2$. As $T$ increases, it is observed in Fig.~\ref{fig:user3_result_t} that the average number of computed task input-bits by the proposed scheme increases significantly over the two benchmark schemes.
This implies the necessity to optimize the time and bandwidth allocations for D2D offloading from the user to the helpers. Due to the benefit of utilizing the shared computation resources of the helpers, it is observed in Fig. \ref{fig:user3_result_t} that the benchmark scheme with uniform beamforming design outperforms the local-computing-only benchmark scheme. In Fig.~\ref{fig:user3_result_p}, similar performance trends are observed as those in Fig.~\ref{fig:user3_result_t}.

\begin{figure}%[!htb]
 \centering
  \includegraphics[width=3.5in]{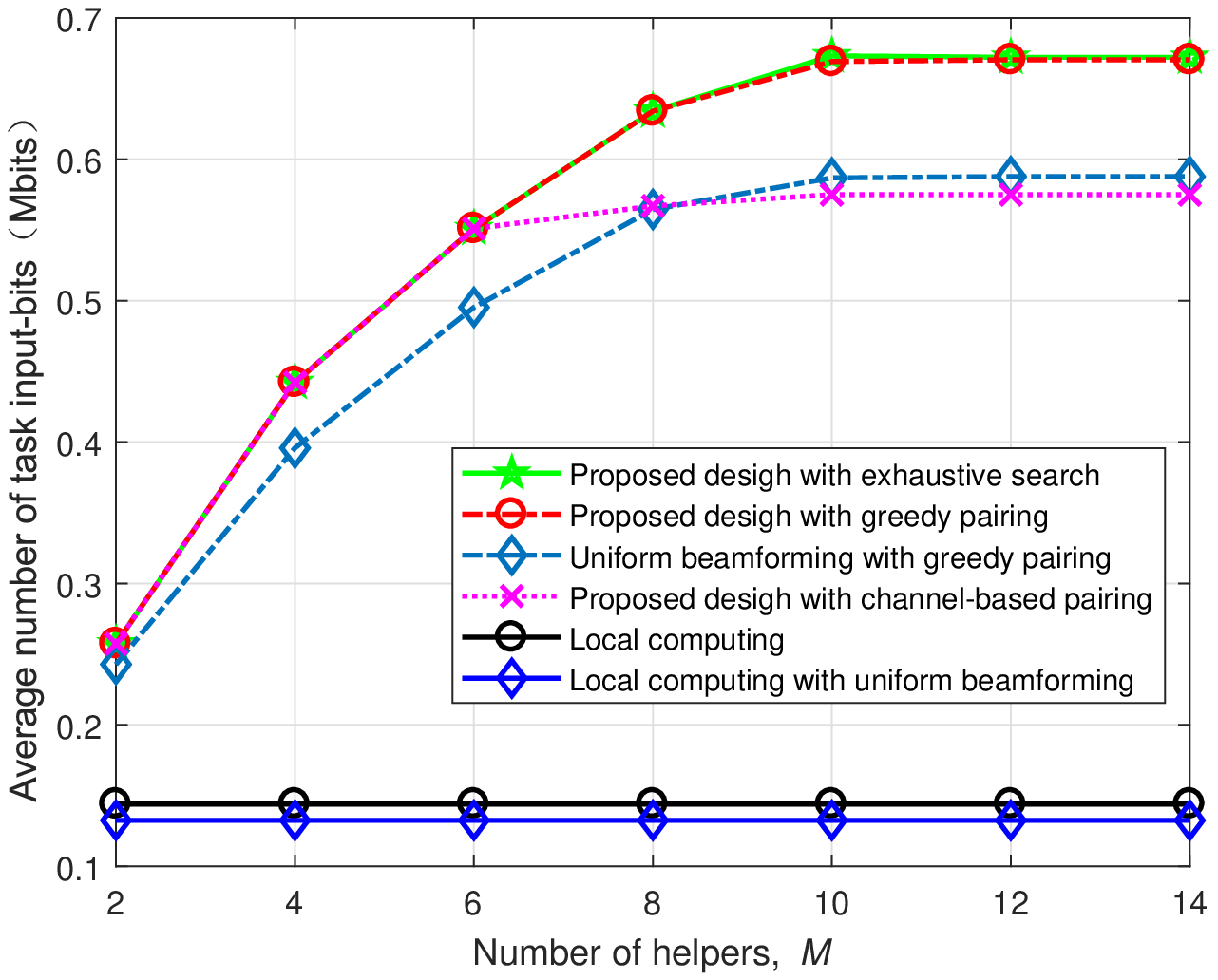}
 \caption{The average number of task input-bits versus the number of helpers.}\label{fig:selection}
\end{figure}

\begin{figure}
\begin{minipage}[t]{0.5\linewidth}
 \includegraphics[width=3.3in]{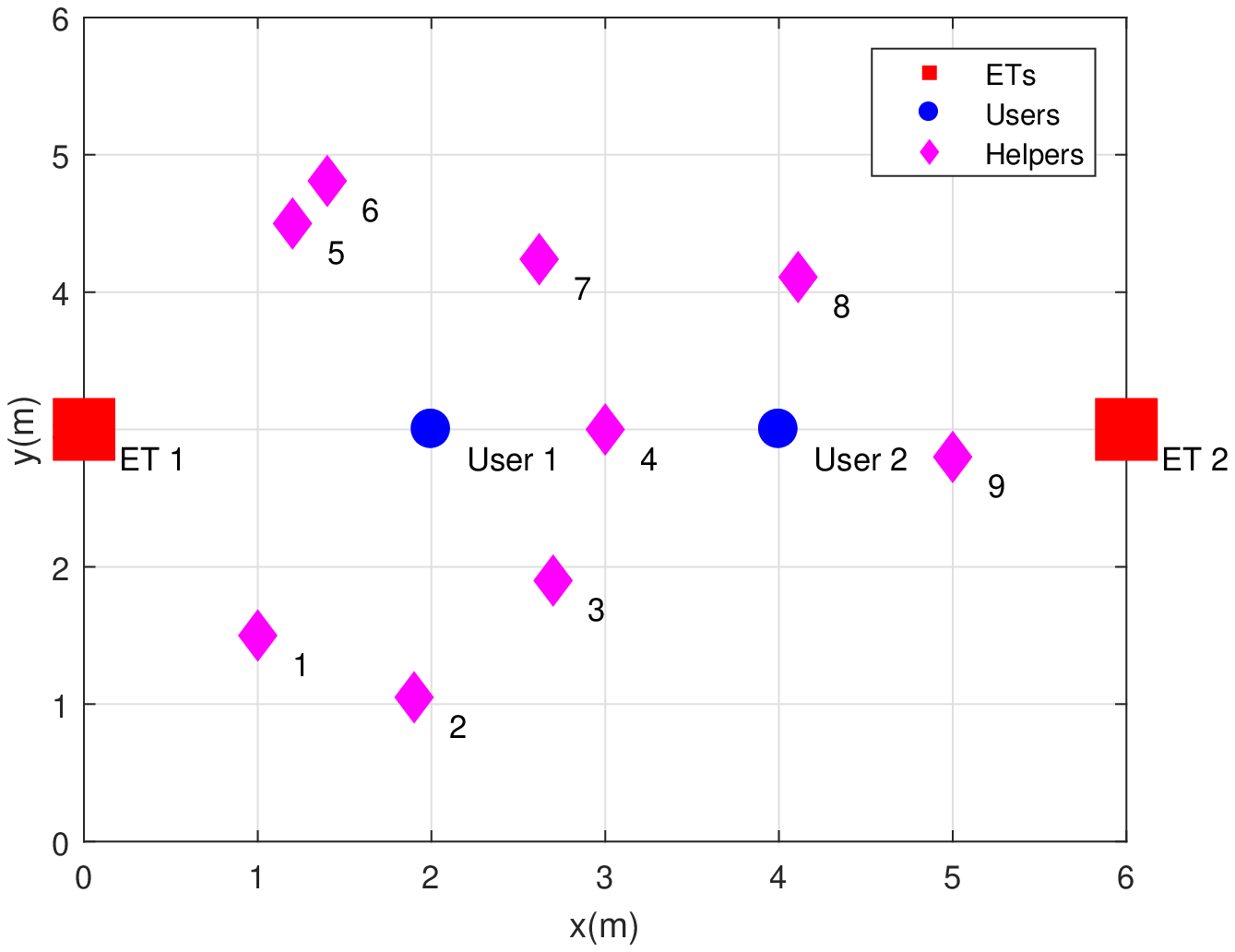}
 \caption{Simulation setup with $N=K=2$ and $M=8$.}\label{fig:near_far_con}
\end{minipage}%
\begin{minipage}[t]{0.5\linewidth}
 \includegraphics[width=3.3in]{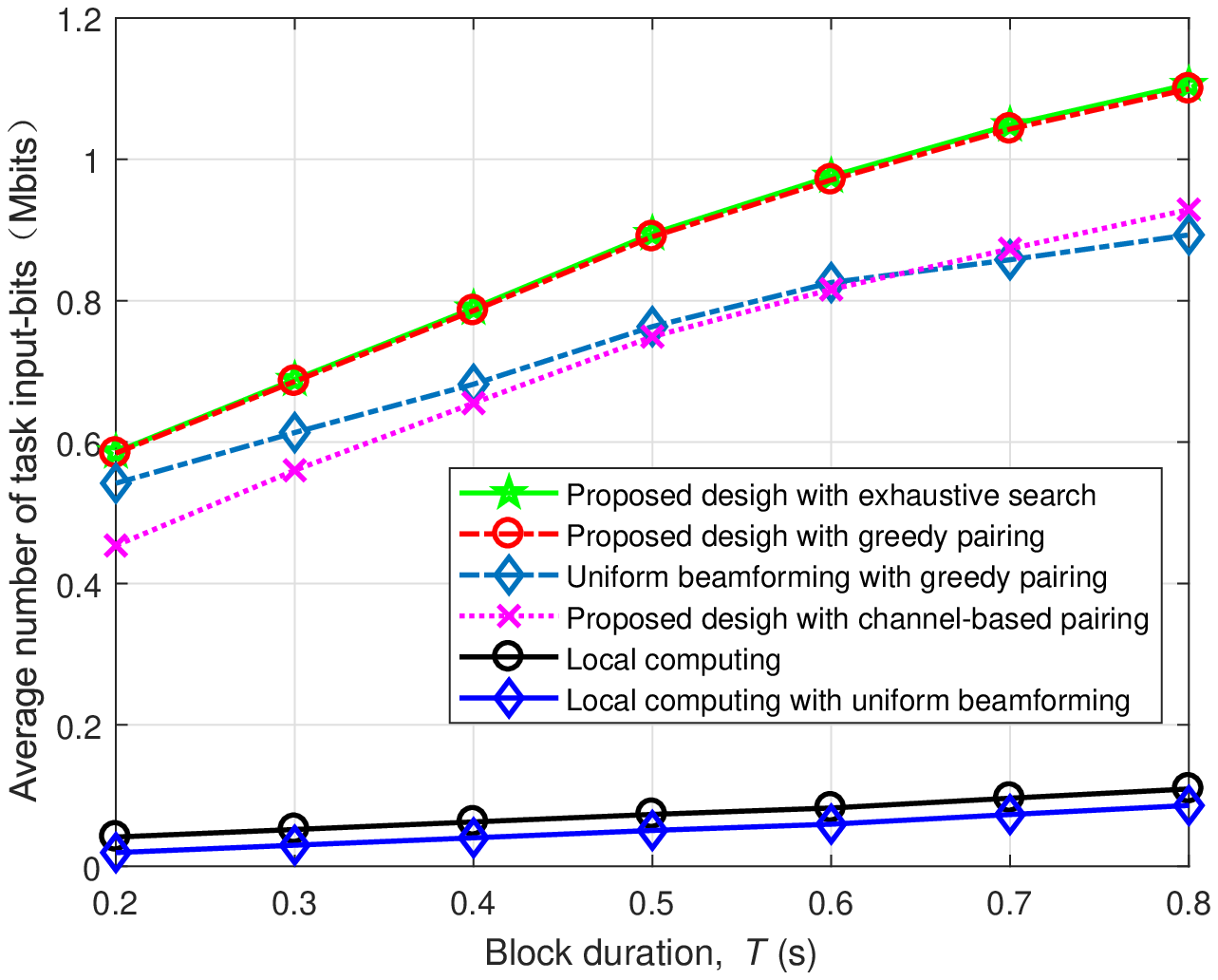}
 \caption{The average number of task input-bits versus the \protect\\block duration $T$.}\label{fig:selection_t}
\end{minipage}
\end{figure}

Next, we consider the multiple-user case with $K=2$ and $N=2$. Fig. \ref{fig:selection} shows the average number of computation task input-bits versus the number of helpers $M$, where the helpers are with different CPU frequencies to execute one CPU cycle, i.e, $f_{k,m} \in \{10^{-26}, 10^{-27},  10^{-28}\}$\cite{13}.
It is observed that the proposed greedy-based pairing scheme achieves a performance very close to that of the optimal exhaustive search method.
It is also observed in Fig. \ref{fig:selection} that the channel-based pairing scheme achieves the near optimal performance at small $M$ values (e.g., $M<6$), but it performs inferior to the greedy-based pairing method scheme as $M$ becomes larger.
When $M$ becomes larger than 6, it is observed that the greedy-based method pairing scheme with uniform beamforming outperforms the channel-based method pairing scheme.
In Fig. 7, the local computing only schemes without or with uniform beamforming are observed to remain unchanged with respect to $M$ and perform inferior to the other four schemes. This further indicates the importance of utilizing the helpers' shared computation resources.

Fig. \ref{fig:near_far_con} presents a wireless powered multiuser D2D offloading setup with $N=2$, $K=2$, and $M=9$. Under this setup, Fig. \ref{fig:selection_t} shows the average number of computation task input-bits versus the block duration $T$. By considering one particular channel realization, we observe that the channel-based pairing scheme allows the helpers $1-7$ to be paired with user $1$. In this case, user $2$ can not sufficiently exploit the nearby helpers' computation resources when compared to user $1$.
By contrast, the proposed greedy-based pairing makes the helpers $1-3, 5-6$ be paired with user $1$ and the remaining helpers paired with user $2$. It is observed in Fig. \ref{fig:selection_t} that the proposed greedy-based pairing scheme achieves close-to-optimal performance of the exhaustive search-based pairing scheme, and significantly outperforms the channel-based pairing scheme.
It is also observed that the local computing design with exhaustive search-based pairing outperforms the joint design with channel-based pairing when $T<0.6$ s, but it is not true when $T$ grows larger in this setup. This implies the importance of designing user-helper pairing to improve the system computation performance.

\section{Conclusion}\label{sec:conclusion}
In this paper, we studied the sum computation rate maximization problem for a wireless powered D2D-enabled cooperative computing system. We jointly optimized the collaborative energy beamforming at the ETs, the communication and computation resource allocations at both the users and helpers, as well as the user-helper pairing sets for performance maximization. Under given user-helper pairing, we leveraged the alternative optimization and convex optimization techniques to optimize the joint communication and computation resource allocation. Furthermore, we designed efficient approaches to pair helpers with users.
Numerical results show that the proposed design significantly improves the computation rate at users, as compared to conventional schemes without such joint optimization. It is expected that this paper can provide new insights on using D2D-enabled offloading and WPT to achieve higher computation performance for low-power wireless IoT devices in a self-sustainable manner, by fully utilizing the energy and computation resources around the mobile edge networks.

\appendix
\subsection{Proof of Lemma \ref{lemma:1}}\label{bounded_appendix}
First, we prove $\bm F(\!\{\!\lambda_{k},\! \mu_{k,m},\! \gamma_n\!\}\!)\!\preceq \!\mv 0$. Assume that $\bm F\!(\!\{\lambda_{k},\! \mu_{k,m},\! \gamma_n\!\}\!)\npreceq \mv 0 $. Denote by $\bm \nu \in {\mathbb C}^{N_tN\times 1}$ an eigenvector corresponding to one positive eigenvalue of $\bm F\left(\{\lambda_{k},\mu_{k,m}, \gamma_n\right\})$. By setting $ \bm S$ = $\tau\bm {\bar \nu}\mv \nu^{\dagger}$ $\geq$ 0 with $\tau$ going to positive infinity, it follows that
\begin{align}
&\lim_{\tau \rightarrow +\infty} { \rm tr}\Big(\bm F\left(\{\lambda_{k},\mu_{k,m}, \gamma_n\right\}) \bm S\Big) = \lim_{\tau \rightarrow +\infty} \tau\mv \nu^{\dagger}\bm F\left(\{\lambda_{k},\mu_{k,m}, \gamma_n\right\})\mv \nu = +\infty,
\end{align}
which in turn implies that the value $g(\{\lambda_{k}, \mu_{k,m}, \rho_{k,m}, \gamma_n\})$ in (D2.1) is unbounded from above over $ \bm S$ $\succeq$ 0. Hence, to ensure that $g(\{\lambda_{k}, \mu_{k,m}, \rho_{k,m}, \gamma_n\})$ is bounded, it requires that constraint
$\bm F(\{\lambda_{k}, \mu_{k,m}, \rho_{k,m}, \gamma_n\})\preceq \bm{0}$.

Note that if $\lambda_k = 0$ or $\mu_{k,m} > 0$, the value of $g(\{\lambda_{k}, \mu_{k,m}, \rho_{k,m}, \gamma_n\})$ becomes positive infinity as $\ell_{k,0}$ or $\ell_{k,m}$ approaches positive infinity. Hence, it follows that $ \lambda_k > 0 $ and $\mu_{k,m} > 0, \forall m\in {\cal M}_k, k\in\cal K$.

\subsection{Proof of Lemma \ref{lem.t_l} }\label{appendix:1}
Given $\{\lambda_{k}, \mu_{k,m}, \rho_{k,m}, \gamma_n\}\in{\cal X}$, we solve problem \eqref{eq.prob_sub3} for each user $k\in{\cal K}$, and then the corresponding Lagrangian is given by
\begin{align}
{\cal L}_m & \triangleq \ell_{k,m}\!-\!\lambda_kt_{k,m}^{(1)} \!\Big(\! 2^{\frac{\ell_{k,m}}{t_{k,m}^{(1)}b_{k,m}}}\!-\!1\!\Big)\! \frac{N_0b_{k,m}}{h_{k,m}}\!-\! \frac{\mu_{k,m}\xi_{k,m}C_{k,m}^3\ell_{k,m}^3}{t_{k,m}^{(2)2}}\!-\!\mu_{k,m} t_{k,m}^{(3)} \!\Big(\!2^{\frac{\beta\ell_{k,m}}{t_{k,m}^{(3)}b_{k,m}}}\!-\!1\!\Big)\! \frac{N_0b_{k,m}}{h_{k,m}}\notag\\
&+\vartheta_m\ell_{k,m}+\sum_{i=1}^3 \Big(\bar{\theta}_{k,m,i}(T-t_{k,m}^{(i)})+\underline{\theta}_{k,m,i}t_{k,m}^{(i)}-\rho_{k,m} t_{k,m}^{(i)}\Big)
\end{align}
where $\vartheta_m$, $\bar{\theta}_{k,m,i}$, denote $\underline{\theta}_{k,m,i}$, are the nonnegative Lagrange multipliers associated with $\ell_{k,m}\geq 0$, $t_{k,m}^{(i)}\leq T$, and $t_{k,m}^{(i)}\geq 0$, respectively. Define $\bar{\bm \theta}_{k,m} \triangleq [\bar{\theta}_{k,m,1},\bar{\theta}_{k,m,2},\bar{\theta}_{k,m,3}]^\dagger$, and $\underline{\bm \theta}_{k,m} \triangleq [\underline{\theta}_{k,m,1},\underline{\theta}_{k,m,2},\underline{\theta}_{k,m,3}]^\dagger$. Based on the KKT conditions, the necessary and sufficient conditions for the optimal primal-dual point $(\ell^*_{k,m},  t_{k,m}^{(i)*},\vartheta_m^*, \bar{\bm \theta}^*_k, \underline{\bm \theta}^*_k)$ are
\begin{subequations}\label{eq.sub_kkt}
	\begin{align}
	&\ell^*_{k,m}\geq 0,~0 \leq t^{(i)*}_{k,m} \leq T, ~~\forall i\in \{1,2,3\}, \\
	&\vartheta_m^*\geq 0,~\bar{\theta}^*_{k,m,i}\geq 0,~\underline{\theta}_{k,m,i}t_{k,m}^{(i)*} \geq 0,~~\forall i\in \{1,2,3\},\\
	&\vartheta_m^*\ell^*_{k,m} = 0,~\bar{\theta}_{k,m,i}(T-t_{k,m}^{(i)})=0,~\underline{\theta}^*_{k,i}t^*_{k,i}=0,~~\forall i\in \{1,2,3\},\\
	&1-\frac{\lambda_kN_0\ln 2}{h_{k,m}}2^{\frac{r^{(1)*}_{k,m}}{N_0b_{k,m}}}-\frac{\beta \mu_{k,m}N_0\ln 2}{ h_{k,m}}2^{\frac{\beta r^{(3)*}_{k,m}}{N_0b_{k,m}}}-{3\mu_{k,m}\xi_{k,m}C_{k,m}^3\left(r^{(i)*}_{k,m}\right)^2}+\vartheta_m^*= 0, \\
	& \frac{\lambda_kN_0b_{k,m}}{h_{k,m}}2^{\frac{r^{(1)*}_{k,m}}{b_{k,m}}}\Big( \frac{r^{(1)*}_{k,m}}{b_{k,m}}\ln2 -1\Big)+\frac{\lambda_kN_0b_{k,m}}{h_{k,m}}-\bar{\theta}^*_{k,m,1}+\underline{\theta}^*_{k,m,1}-\rho_{k,m} = 0,\\
	& 2\mu_{k,m}\xi_{k,m}C_{k,m}^3 r^{(2)*}_{k,m} -\bar{\theta}^*_{k,m,2}+\underline{\theta}^*_{k,m,2} -\rho_{k,m}= 0,\\
	& \frac{\mu_{k,m}N_0b_{k,m}}{h_{k,m}}2^{\frac{\beta r^{(3)*}_{k,m}}{b_{k,m}}}\Big(\frac{\beta r^{(3)*}_{k,m}}{b_{k,m}}\ln2-1\Big)+\frac{\mu_{k,m}N_0b_{k,m}}{h_{k,m}}-\bar{\theta}^*_{k,m,3}+
\underline{\theta}^*_{k,m,3}-\rho_{k,m}=0,
	\end{align}
\end{subequations}
where $r^{(i)*}_{k,m} \triangleq \ell_{k,m}^*/t_{k,m}^{(i)*}$, $i\in \{1,2,3\}$. The left-hand-side (LHS) terms of (\ref{eq.sub_kkt}d)--(\ref{eq.sub_kkt}g) are the first-order derivatives of ${\cal L}_m$ with respect to $\ell^*_{k,m}$, $t_{k,m}^{(2)*}$, $t_{k,m}^{(2)*}$, and $t_{k,m}^{(3)*}$, respectively. From (\ref{eq.sub_kkt}b), (\ref{eq.sub_kkt}c), and (\ref{eq.sub_kkt}e), it follows that
\begin{align}\label{eq.kkt_r1}
\left(\frac{r^{(1)*}_{k,m}}{b_{k,m}}\ln2-1\right)e^{\left(\frac{r^{(1)*}_{k,m}}{b_{k,m}}\ln2-1\right)} = \frac{\rho_{k,m}h_{k,m}}{\lambda_kN_0b_{k,m} e}-\frac{1}{e}.
\end{align}
For the function $y=xe^x$ of $x>0$, its inverse function can be shown to be $x=W(y)$ \cite{18}. Therefore, based on \eqref{eq.kkt_r1} and algebra manipulation, we have
\begin{align}
r^{(1)*}_{k,m}= \frac{b_{k,m}}{\ln2}\Big(1+W\Big(\frac{\rho_{k,m}h_{k,m}}{\lambda_kN_0b_{k,m}e}-\frac{1}{e}\Big)\Big).
\end{align}
Based on (\ref{eq.sub_kkt}b) and (\ref{eq.sub_kkt}f), it follows that
\begin{align}
r^{(2)*}_{k,m}= \frac{1}{C_{k,m}}\left(\frac{\rho_{k,m}}{2\mu_{k,m}\xi_{k,m}}\right)^{\frac{1}{3}}.
\end{align}
Similarly, from (\ref{eq.sub_kkt}b) and (\ref{eq.sub_kkt}g), we have
\begin{align}
r^{(3)*}_{k,m}= \frac{b_{k,m}}{\beta\ln2}\Big(1+W\Big(\frac{\rho_{k,m}h_{k,m}}{\mu_{k,m}N_0b_{k,m}e}-\frac{1}{e}\Big)\Big).
\end{align}
%With $r^*_{k,i}$, $i\in{\cal I}$, we next obtain the optimal $\ell_k$ based on (\ref{eq.sub_kkt}c) and (\ref{eq.sub_kkt}d). Specifically, from (\ref{eq.sub_kkt}d), it follows that
%\begin{align}
%\gamma^*_k = 1-\frac{\lambda_0\sigma_{{\rm{u}},m}^2\ln 2}{Bh_{k,m}}2^{\frac{r_{k,1}^*}{B}}-\frac{\beta \lambda_k\sigma_0^2\ln 2}{B h_{k,m}}2^{\frac{\beta r^*_{k,3}}{B}}-{3\lambda_k\xi_kC_k^3 r_{k,i}^{*2}}.
%\end{align}
%Together with the complementary slackness constraint $\gamma_k^*\ell_k^*=0$, we have
%\begin{align}
%\ell^*_k
%\begin{cases}
%=0,& {\tt if}~\gamma^*_k>0\\
%\geq 0,& {\tt if} ~\gamma_k^* = 0
%\end{cases}
%\end{align}
To determine the optimal $\ell^*_{k,m}$ to problem \eqref{eq.prob_sub3}, we re-express problem \eqref{eq.prob_sub3} as
	\begin{align}\label{prob.ell_m}
	\max_{\ell_{k,m}}~&\ell_{k,m} G\left(r^{(i)*}_{k,m},\lambda_k, \mu_{k,m},\rho_{k,m}\right)\\
	{\text{ s.t.}}~ ~&0 \leq \ell_{k,m} \leq r^{(i)*}_{k,m}T,~\forall i\in\{1,2,3\},\notag
	\end{align}
where $G\left(r^{(i)*}_{k,m},\lambda_k, \mu_{k,m},\rho_{k,m}\right)$ is given in \eqref{G_r}.
%$G\left(r^{(i)*}_{k,m},\lambda_k, \mu_{k,m},\rho_{k,m}\right) \triangleq 1-\frac{\lambda_kN_0b_{k,m}}{h_{k,m}r^{(1)*}_{k,m}}\left(2^{\frac{r^{(1)*}_{k,m}}{b_{k,m}}}-1\right)-\sum_{i=1}^3 \frac{\rho_{k,m}}{r^{(1)*}_{k,m}}-\frac{\lambda_kN_0b_{k,m}}{h_{k,m} r^{(3)*}_{k,m}}\left(2^{\frac{\beta r^{(3)*}_{k,m}}{b_{k,m}}}-1\right)-\mu_{k,m}{\xi_{k,m}C_{k,m}^3\left(r^{(3)*}_{k,m}\right)^2},~ k\in{\cal K}$.
It readily follows that the optimal $\ell_{k,m}^*$ to problem \eqref{prob.ell_m} is given by
\begin{align*}
\ell_{k,m}^*
\begin{cases}
=0,~&~{\rm if}~G\left(\rm r^{(i)*}_{k,m},\lambda_k, \mu_{k,m},\rho_{k,m}\right)<0\\
\in \left[0,\min_{i\in\{1,2,3\}} r^{(i)*}_{k,m}T \right],~&~{\rm if}~G\left(r^{(i)*}_{k,m},\lambda_k, \mu_{k,m},\rho_{k,m}\right)=0\\
=\min_{i\in\{1,2,3\}}r^{(i)*}_{k,m}T, ~&~{\rm if}~ G\left(r^{(i)*}_{k,m},\lambda_k, \mu_{k,m},\rho_{k,m}\right)> 0.
\end{cases}
\end{align*}
Then, we have $t^{(i)*}_{k,m} = \frac{\ell_{k,m}}{r^{(i)*}_{k,m}}$ and Lemma 4 is proved finally.

\subsection{Proof of Lemma \ref{lem.sub.F}}
The negative semidefinite constraint  $\bm F\left(\{\lambda_{k},\mu_{k,m}, \gamma_n\right\})$ $\preceq \mv 0$ can be equivalently expressed as a scalar inequality constraint as
\begin{align}
\pi(\{\lambda_{k}, \mu_{k,m}, \gamma_n\})\triangleq \max_{||\mv \varrho||=1} \mv\varrho^{\dagger}F(\{\lambda_{k}, \mu_{k,m}, \rho_{k,m}, \gamma_n\})\mv\varrho\leq 0.
\end{align}
Suppose there is a point $(\{\lambda_{k}, \mu_{k,m}, \gamma_n\}) \triangleq [\lambda_{1,1},\ldots,\lambda_{1,K},\mu_{1,1},\ldots,\mu_{1,M},\gamma_{1,1},\ldots,\gamma_{1,N}]^\dagger$, and one can find the normalized eigenvector $\bm \nu_{1}$ of $\bm F(\{\lambda_{k}, \mu_{k,m}, \gamma_n\})$ corresponding to the maximal eigenvalue of $\mv F(\{\lambda_{k}, \mu_{k,m}, \gamma_n\})$, which means $\pi(\{\lambda_{k}, \mu_{k,m}, \gamma_n\})=\bm \nu^{\dagger}\bm{F}(\{\lambda_{k}, \mu_{k,m}, \gamma_n\})\mv\nu$. The subgradient can be expressed as
\begin{align}
&\pi(\{\lambda_{k}, \mu_{k,m}, \gamma_n\})-\pi(\{\lambda_{k}, \mu_{k,m}, \gamma_n\})\notag\\
=& \max_{||\mv \varrho||=1} \mv\varrho^{\dagger}\bm{F}\left(\{\lambda_{k},\mu_{k,m}, \gamma_n\right\})\mv\varrho-\mv\nu^{\dagger}\bm{F}(\{\lambda_{k}, \mu_{k,m}, \gamma_n\})\mv\nu\notag\\
\leq& \mv\nu^{\dagger}\bm{F}(\left(\{\lambda_{k},\mu_{k,m}, \gamma_n\right\})-\bm{F}(\{\lambda_{k}, \mu_{k,m}, \gamma_n\}))\mv\nu\notag\\
=&\sum_{k \in \cal K} (\lambda_{1,k}-\lambda_k)T\eta{\bm g}_{k,0}{\bm g}_{k,0}^{\dagger}+\sum_{n \in \cal N}(\gamma_{1,n}-\gamma_n)\bm \Sigma_n\notag\\
+&\sum_{m \in {\cal M}_k}(\mu_{1,m}-\mu_{k,m})T\eta{\bm g}_{k,m}{\bm g}_{k,m}^{\dagger},
\end{align}
where the last equality follows from the affine structure of $\bm F(\cdot)$ in problem (D2.1). The subgradient of $\bm F\left(\{\lambda_{k},\mu_{k,m}, \gamma_n\right\})$ at the given $\big[T\eta\mv \nu^{H}{\bm g}_{1,0}{\bm g}_{1,0}^{H}\mv \nu,\ldots,T\eta\mv \nu^{H}{\bm g}_{K,0}{\bm g}_{K,0}^{H}\mv \nu, T\eta\mv \nu^{H}\bm g_{k,1}{\bm g}_{k,1}^{H}\mv \nu, \ldots,\\ T\eta\mv \nu^{H}\bm g_{k,M}{\bm g}_{k,M}^{H}\mv \nu, 0, \ldots, 0, \mv \nu^{H} \bm \Sigma_1\mv \nu,\ldots,{\mv \nu^{H}} \bm \Sigma_N{\mv \nu} \big]^{\dagger}$, where $\bm \nu$ is the eigenvector corresponding to the maximal eigenvalue of $\bm F\left(\{\lambda_{k},\mu_{k,m}, \gamma_n\right\})$.

%\ifCLASSOPTIONcaptionsoff
%  \newpage
%\fi

%\bibliography{mec_May27.bib}
%\bibliographystyle{IEEEtran}
%\bibliography{mec_May27.bib}

\begin{thebibliography}{1}

\bibliographystyle{IEEEbib}

\bibitem{Conf_version}
D. Wu, F. Wang, X. Cao, and J. Xu, ``Wireless powered user cooperative computation in mobile edge computing systems," in {\it Proc. IEEE Globecom Workshops}, Abu Dhabi, United Arab Emirates, 2018, pp. 1--7.

\bibitem{1}
S. Barbarossa, S. Sardellitti, and P. D. Lorenzo, ``Communicating while computing: Distributed mobile cloud computing over 5G heterogeneous networks," {\it IEEE Signal Process. Mag.}, vol. 31, no. 6, pp. 45--55, Nov. 2014.

\bibitem{2}
Y. Hu, M. Patel, D. Sabella, N. Sprecher, and V. Young, ``Mobile edge computing: A key technology towards 5G," ETSI, Sophia Antipolis, France, White Paper 11, 2015. [Online]. Available: \url{http://www.etsi.org/images/files/ETSIWhitePapers/etsi/5g.pdf}
%\bibitem{2}
%E. Cuervo, A. Balasubramanian, D. Cho, A. Wolman, S. Saroiu, R.Chandra, and P. Bahl, ``MAUI: Making smartphones last longer with code offload," {\it Proc. ACM MobiSys.}, San Francisco, USA, Jun. 2010, pp. 49--62.
%\bibitem{3}
%S. Kosta, A. Aucinas, P. Hui, R. Mortier, and X. Zhang, ``ThinkAir: Dynamic resource allocation and parallel execution in the cloud for mobile code offloading," {\it Proc. IEEE INFOCOM}, Orlando, USA, Mar. 2012, pp. 945--953.
\bibitem{3}	
Y. Mao, C. You, J. Zhang, K. Huang, and K. B. Letaief, ``A survey on mobile edge computing: The communication perspective," {\it IEEE Commun. Survey Tuts.}, vol. 19, no. 4, pp. 2322--2358, 4th Quart. 2017.
%\bibitem{99}	
%M. H. Chen, B. Liang, and M. Dong, ``Joint offloading decision and resource allocation for multi-user multi-task mobile cloud," in  {\em Proc. IEEE ICC}, Kuala, Lumpur, 2016, pp. 1--6.
%\bibitem{1111}
%S. Yu, R. Langar, and X. Wang, ``A D2D-multicast based computation offloading framework for interactive applications," in {\it Proc. IEEE Globecom}, Washington, DC, 2016, pp. 1--6.
%\bibitem{6}
%J. Liu, Y. Mao, J. Zhang and K. B. Letaief, ``Delay-optimal computation task scheduling for mobile-edge computing systems," {\it Proc. IEEE ISIT}, Barcelona, 2016, pp. 1451--1455.

%\bibitem{8}
%W. Zhang, Y. Wen, K. Guan, D. Kilper, H. Luo and D. O. Wu, ``Energy-optimal mobile cloud computing under stochastic wireless channel," {\it IEEE Trans. Wireless Commun.}, vol. 12, no. 9, pp. 4569--4581, Sep. 2013.

\bibitem{4}
J. Xu and R. Zhang, ``Energy beamforming with one-bit feedback," {\it IEEE Trans. Signal Process.}, vol. 62, no. 20, pp. 5370--5381, Oct. 2014.
\bibitem{general}
J. Xu and R. Zhang, ``A general design framework for MIMO wireless energy transfer with limited feedback," {\it IEEE Trans. Signal Process.},vol. 64, no. 10, pp. 2475--2488, May 2016.
\bibitem{5}
J. Xu, Y. Zeng, and R. Zhang, ``UAV-enabled wireless power transfer: Trajectory design and energy optimization," {\it IEEE Trans. Wireless Commun.}, vol. 17, no. 8, pp. 5092--5106, Aug. 2018.
\bibitem{6}
Y. Huang and B. Clerckx, ``Waveform design for wireless power transfer with limited feedback," {\it IEEE Trans. Wireless Commun.}, vol. 17, no. 1, pp. 415--429, Jan. 2018.
\bibitem{7}
Y. Zeng, B. Clerckx, and R. Zhang, ``Communications and signals design for wireless power transmission," {\it IEEE Trans. Commun.}, vol. 65, no. 5, pp. 2264--2290, May 2017.

%\bibitem{77}
%R. Atat, L. Liu, N. Mastronarde, and Y. Yi, ``Energy harvesting based D2D-assisted machine-type communications," {\it IEEE Trans. Commun.}, vol. 65, no. 3, pp. 1289--1302, Mar. 2017.
%\bibitem{88}
%H. H. Yang, J. Lee, and T. Q. S. Quek, ``Heterogeneous cellular network with energy harvesting-based D2D communication," {\it IEEE Trans. Wireless Commun.}, vol. 15, no. 2, pp. 1406--1419, Feb. 2016
\bibitem{8}
J. Xu, L. Liu, and R. Zhang, ``Multiuser MISO beamforming for simultaneous wireless information and power transfer," {\it IEEE Trans. Signal Process.}, vol. 62, no. 3, pp. 4798--4810, Sep. 2014.

\bibitem{10}
X. Lu, P. Wang, D. Niyato, D. I. Kim, and Z. Han, ``Wireless networks with RF energy harvesting: A contemporary survey," {\it IEEE Commun. Surveys Tuts.}, vol. 17, no. 2, pp. 757--789, 2nd Quart. 2015.

\bibitem{lifeng}
L. Xie, J. Xu, and R. Zhang, ``Throughput maximization for UAV-enabled wireless powered communication networks,'' {\it IEEE Internet Things J.}, vol. 6, no. 2, pp. 1690--1703, Apr. 2019.

\bibitem{bisuzhi}
S. Bi, C. K. Ho, and R. Zhang, ``Wireless powered communication: Opportunities and challenges," {\it IEEE Commun. Mag.},  vol. 53, no. 4, pp. 117--125, Apr. 2015.

\bibitem{khuang}
K. Huang and V. K. N. Lau, ``Enabling wireless power transfer in cellular networks: Architecture, modeling and deployment," {\it IEEE Trans. Wireless Commun.},  vol. 13, no. 2, pp. 902--912, Feb. 2014.

\bibitem{mxia}
M. Xia and S. Aissa, ``On the efficiency of far-field wireless power transfer," {\it IEEE Trans. Signal Process.},  vol. 63, no. 11, pp. 2835--2847, Jun. 2015.

\bibitem{12}
C. You, K. Huang, and H. Chae, ``Energy efficient mobile cloud computing powered by wireless energy transfer," {\it IEEE J. Sel. Areas Commun.}, vol. 34, no. 5, pp. 1757--1771, May 2016.

\bibitem{13}
F. Wang, J. Xu, X. Wang, and S. Cui, ``Joint offloading and computing optimization in wireless powered mobile-edge computing systems," {\it IEEE Trans. Wireless Commun.}, vol. 17, no. 3, pp. 1784--1797, Mar. 2018.

\bibitem{14}
S. Bi and Y. J. A. Zhang, ``Computation rate maximization for wireless powered mobile-edge computing with binary computation offloading," {\it IEEE Trans. Wireless Commun.}, vol. 17, no. 6, pp. 4177--4190, Jun. 2018.
%\bibitem{bshang}
%B. Shang, L. Zhao, K.-C. Chen, and X. Chu, ``Energy efficient D2D assisted offloading with wireless power transfer," in {\em Proc. IEEE GLOBECOM}, Singapore, Dec. 2017, pp. 1--6.

\bibitem{15}
X. Hu, K.-K. Wong, and K. Yang, ``Wireless powered cooperation-assisted mobile edge computing," {\it IEEE Trans. Wireless Commun.}, vol. 17, no. 4, pp. 2375--2388, Apr. 2018.
\bibitem{wuqingqing}

Q. Wu, G. Y. Li, W. Chen, and D. W. K. Ng, ``Energy-efficient D2D overlaying communications with spectrum-power trading," {\it IEEE Trans. Wireless Commun.}, vol. 16, no. 7, pp. 4404--4419, Jul. 2017.
\bibitem{fengdaquan}

D. Feng, L. Lu, Y. Yuan-Wu, G. Y. Li, G. Feng, and S. Li, ``Device-to-device
communications underlaying cellular networks," {\it IEEE Trans. Commun.}, vol. 61, no. 8, pp. 3541--3551, Aug. 2013.

\bibitem{xiaowen}
X. Cao, F. Wang, J. Xu, R. Zhang, and S. Cui, ``Joint computation and communication cooperation for energy-efficient mobile edge computing," {\it IEEE Internet Things J.}, vol. 6, no. 3, pp. 4188--4200, Jun. 2019.

\bibitem{qijieLin}
Q. Lin, F. Wang, and J. Xu, ``Optimal task offloading scheduling for energy efficient D2D cooperative computing," {\it IEEE Commun. Lett.}, vol. 23, no. 10, pp. 1816--1820, Oct. 2019.

\bibitem{hongXing}
H. Xing, L. Liu, J. Xu, and A. Nallanathan, ``Joint task assignment and resource allocation for D2D-enabled mobile-edge computing," {\it IEEE Trans. Commun.}, vol. 67, no. 6, pp. 4193--4207, Jun. 2019.
\bibitem{pulingjun}

L. Pu, X. Chen, J. Xu, and X. Fu, ``D2D fogging: An energy-efficient and incentive-aware task offloading framework via network-assisted D2D collaboration," {\it IEEE J. Sel. Areas Commun.}, vol. 34, no. 12, pp. 3887--3901, Nov. 2016.

\bibitem{chenxu}
X. Chen, L. Pu, L. Gao, W. Wu, and D. Wu, ``Exploiting massive D2D collaboration for energy-efficient mobile edge computing," {\it IEEE Wireless Commun.},  vol. 24, no. 4, pp. 64--71, Aug. 2017.
%\bibitem{12}
%J. Xu and R. Zhang, ``Energy beamforming with one-bit feedback," {\it IEEE Trans. Signal Process.}, vol. 62, no. 20, pp. 5370--5381, Oct. 2014.
%\bibitem{13}
%J. Xu and R. Zhang, ``A general design framework for MIMO wireless energy transfer with limited feedback," {\it IEEE Trans. Signal Process.},vol. 64, no. 10, pp. 2475--2488, May 2016.
%\bibitem{14}
%C. Valenta and G. Durgin, ``Harvesting wireless power: Survey of energy-harvester conversion efficiency in far-field, wireless power transfer systems," {\it IEEE Microw. Mag.}, vol. 15, no. 4, pp. 108--120, Jun. 2014.

\bibitem{18}
R. Corless, G. Gonnet, D. Hare, D. Jeffrey, and D. Knuth, ``On the Lambert $W$ function," {\it Adv. Comput. Math.}, vol. 5, no. 1, pp. 329--359, Dec. 1996.

\bibitem{votage}
T. D. Burd and R. W. Brodersen, ``Processor design for portable systems," {\it J. VLSI Signal Process. Syst.}, vol. 13, no. 2, pp. 203--221, Aug. 1996.

\bibitem{19}\label{cvx}
S. Boyd and L. Vandenberghe, {\it Convex Optimization}, Cambridge, U.K.: Cambridge Univ. Press, Mar. 2004.

\bibitem{20}
S. Boyd, ``Ellipsoid method,"  Stanford Univ., Stanford, CA, USA, May 2014.  [Online]. Available: \url{https://web.stanfordedu/class/ee364b/lectures/ellipsoid method slides.pdf}
\end{thebibliography}
%\bibliographystyle{IEEEtran}

\end{document}